\documentclass[numberedappendix,twocolumn,onecolappendix]{openjournal}

\usepackage{newtxtext,newtxmath}
\usepackage{fdsymbol}
\usepackage{longtable}

\usepackage[T1]{fontenc}

\DeclareRobustCommand{\VAN}[3]{#2}
\let\VANthebibliography\thebibliography
\def\thebibliography{\DeclareRobustCommand{\VAN}[3]{##3}\VANthebibliography}

\usepackage{graphicx}	
\usepackage{amsmath}	
\usepackage{amssymb}

\usepackage{lipsum}
\usepackage[dvipsnames]{xcolor}
\usepackage{subfigure}
\usepackage{graphicx}
\usepackage{hyperref}	
\hypersetup{colorlinks=true,linkcolor=blue,citecolor=blue,filecolor=blue,urlcolor=blue}
\usepackage[export]{adjustbox} 
\usepackage{booktabs}

\newcommand{\Msun}{M_\odot}

\begin{document}
\title[Machine Learning Halo Properties II]{Machine Learning the Dark Matter Halo Mass of the Milky Way}

\author{Elaheh Hayati$^{1,*,\dag}$}

\author{Peter Behroozi$^{1}$}

\author{Ekta Patel$^{2,3,\ddag}$}

\author{Yunchong Wang}

\author{Stefan Gottl{\"o}ber$^{5}$}

\author{Gustavo Yepes$^{6,7}$}

\affiliation{$^{1}$Department of Astronomy and Steward Observatory, University of Arizona, 933 N Cherry Ave, Tucson, AZ, 85721, USA}


\affiliation{$^{2}$Department of Astrophysics and Planetary Sciences, Villanova University,  800 E. Lancaster Ave, Villanova, PA 19085, USA}

\affiliation{$^{3}$Department of Physics and Astronomy, University of Utah, 115 South 1400 East, Salt Lake City, Utah 84112, USA}

\affiliation{$^{5}$ Leibniz-Institut f\"ur Astrophysik Potsdam (AIP), An der Sternwarte 16, D - 14482, Potsdam, Germany}

\affiliation{$^{6}$ Departamento de F\'{\i}sica Te\'orica M-8, Universidad Aut\'onoma de Madrid, Cantoblanco, E-28049 Madrid, Spain }   
        
\affiliation{$^{7}$ Centro de Investigaci\'on Avanzada en F\'{\i}sica  Fundamental (CIAFF), Universidad Aut\'onoma de Madrid, E-28049 Madrid, Spain}

\thanks{* E-mail: ehayati@arizona.edu}
\thanks{$\dag$ LSSTC DSFP Fellow}
\thanks{$\ddag$ Hubble Fellow}

\shortauthors{Hayati et al.}



\begin{abstract}
Although numerous dynamical techniques have been developed to estimate the total dark matter halo mass of the Milky Way, it remains poorly constrained, with typical systematic uncertainties of 0.3 dex. 
In this study, we apply a neural network-based approach that achieves high mass precision without several limitations that have affected past approaches; for example, we do not assume dynamical equilibrium, nor do we assume that neighboring galaxies are bound satellites. Additionally, this method works for a broad mass range, including for halos that differ significantly from the Milky Way. Our model relies solely on observable dynamical quantities, such as satellite orbits, distances to larger nearby halos, and the maximum circular velocity of the most massive satellite. In this paper, we measure the halo mass of the Milky Way to be $\log_{10} M_\mathrm{vir}/\Msun = 12.20^{+0.163}_{-0.138}$. Future studies in this series will extend this methodology to estimate the dark matter halo mass of M31, and develop new neural networks to infer additional halo properties including concentration, assembly history, and spin axis.
\end{abstract}




\section{Introduction}
\label{s:intro}

Within the $\Lambda$CDM framework, the Milky Way is surrounded by a dark matter halo accounting for most of its total mass. However, as dark matter cannot be observed directly, measuring this mass around the Milky Way (MW) has proven challenging. Consequently, numerous studies have sought to estimate the dark matter content of the Milky Way through indirect methods \citep[e.g.,][]{Oort26, Morrison00,Yanny00,Battaglia5, Frinchaboy8,Li08,Busha11, vanderMarel12,king15,Lowing15,Patel17,McMillan17,Patel18,2023arXiv230605461J,2024MNRAS.528..693O,2024ApJ...972...70R,2025A&A...693L...8L}. 

These methods can be broadly divided into two categories: dynamical equilibrium approaches, which include rotation curves and modeling tracers, and simulation-based techniques, such as modeling the motion of the Milky Way and Andromeda as well as finding halos with similar satellite properties (see the review in \citealt{Wang20} and references therein).  Each category has its own advantages and disadvantages. For example, equilibrium approaches offer a simple way to combine arbitrary tracer particles (e.g., rotation curves, globular clusters, halo stars, satellites, etc.), but result in hard-to-quantify biases due to the lack of equilibrium in the Milky Way (e.g., from the passage of the Large Magellanic Cloud and continued mass accretion onto the outer halo; \citep[][]{2020MNRAS.498.5574E,2021ApJ...919..109G,2024OJAp....7E..50K}
.  Past simulation-based approaches result in better treatment of non-equilibrium dynamics, but with a cost that increases exponentially with the number of observables that need to be matched simultaneously.

Two recent papers have developed a third category of approaches, relying on machine learning to approximate the correct mass distribution as a function of an arbitrary combination of observables.  \cite{VD23} used hydrodynamical simulations with inputs of the stellar masses, positions, and velocities of satellite halos to estimate the mass of the Milky Way; however, this paper assumed knowledge of satellite status and did not include observational uncertainties.  More recently, in \citet[][]{2024OJAp....7E..74H}, we developed a neural network method that has many appealing features for measuring halo masses.  Our method does not rely on assumptions of dynamical equilibrium, nor does it assume that nearby galaxies are predominantly satellites. Further, arbitrary local or broader environmental constraints (such as distance or velocity differences to the nearest larger halo) can be incorporated self-consistently.  Last, the technique is capable of inferring relationships between observable phenomena (such as satellite trajectories) and mass, even from halos that do not resemble the MW or M31.

In this paper, we extend our previous method by developing a realistic model for uncertainties and selection effects associated with observations of satellites around the Milky Way.  This allows us to construct mock observations from simulated dark matter halos to realistically model what we would expect to see from \textit{Gaia} DR3 and other observations.  We then train neural networks on synthetic DR3 observations and apply them to the actual data to estimate the mass of the Milky Way's dark matter halo. Unlike \cite{VD23}, we do not include the stellar mass of the Milky Way as an input.  While stellar masses can help constrain the halo mass,  stellar mass measurements (as well the inferred galaxy--halo connection) are only understood at the $\sim 0.3$ dex level \citep[see][for reviews]{Conroy13,Wechsler18}, which could result in biased halo mass inferences. Instead, we combine nearby galaxy orbital information (distance and 3D velocity), dynamical information from the largest satellite, and distances to the nearest larger halo and the nearest larger cluster.


This paper is divided as follows: in Section \ref{s:methods}, we outline the training process and describe the dark matter simulations used in our analysis; in Section \ref{s:data} we discuss the observational selection effects; Section \ref{s:results} presents the Milky Way mass estimates and uncertainties from the trained neural networks, followed by  discussion and conclusions in Section \ref{s:discussion}.  We adopt a flat $\Lambda$CDM cosmology with $\Omega_{m} = 0.307$, $\Omega_{\Lambda} = 0.693$, $n_{s}=0.96$, $h=0.68$ and $\sigma_{8} =0.823$, consistent with \cite{Planck16}.
We adopt the definition of virial halo mass ($M_\mathrm{vir}$) from \citep{Bryan98}, representing the total mass (both dark matter and baryonic matter) enclosed within a radius $R_\mathrm{vir}$ around a density peak.  Where appropriate, we convert $M_{\rm vir}$ to $M_{200c}$ (average halo density corresponding to 200 times the critical density) by subtracting an offset of 0.063 dex. We measured this offset from Rockstar halo catalogs at $z=0$ near $M_{\rm vir}\sim10^{12}\,M_\odot$ in ESMDPL/VSMDPL. This value is consistent with the expected small mass and cosmology dependent variations reported by \citet[][]{2021MNRAS.500.5056R}.

\section{Methods}

\label{s:methods}

\subsection{Dark Matter Simulations}

In this study, we utilize the publicly available Very Small MultiDark Planck (VSMDPL) simulation, which includes $3840^{3}$ dark matter particles, each with a mass of $6.2 \times 10^{6}\Msun/h$, inside a periodic cube with a side length of 160 comoving Mpc/$h$ \citep{Klypin16,RP16}. VSMDPL marginally resolves halos with $M_\mathrm{vir}<10^9\Msun$, which are expected to host satellites fainter than the classical dwarfs, so we also use the Extremely small MultiDark Planck (ESMDPL) simulation (\textit{ibid}), which consists of $4096^{3}$ dark matter particles, each with a mass of $3.25 \times 10^{5}\Msun/h$, in a box size of 64 Mpc/$h$ \citep[][]{2020yCat.6146....0O}. These simulations are conducted within a flat, $\Lambda$CDM universe, with cosmology given as in Section \ref{s:intro}, and they track the evolution of matter from $z=150$ to $z=0$. The VSMDPL simulation contains 151 snapshots, and the ESMDPL simulation contains 70 snapshots.  In both simulations, halos are identified over $z=0-25$, using the  \textsc{Rockstar} \citep{Behroozi013} halo finder, with merger trees from the \textsc{Consistent Trees} code \citep{Behroozi2013}. Each halo in the merger trees is classified either as a central halo or a satellite halo (i.e., a halo located within the virial radius of a larger halo). 

\subsection{The Galaxy--Halo Connection}

N-body simulations do not fully capture the dynamics of satellite galaxies because a satellite galaxy may 1) disrupt earlier or later than the satellite's halo, or 2) have a modified orbital evolution due to baryonic effects.  The first effect can be accounted for by comparing N-body simulation results to the observed number counts of satellites around Milky Way-like galaxies, and adjusting satellite halo lifetimes accordingly. In practice, this correction reflects both genuine physical disruption (e.g., due to baryonic tides or dynamical friction) and numerical disruption in the simulations, so that the total number of surviving satellite galaxies and their radial distributions match observations \citep[][]{2024ApJ...970..178M,2025arXiv250713448W}. Capturing changes to satellite orbits due to baryonic effects is a second-order effect that is beyond the scope of this paper, as it involves cross-training on hydrodynamical simulations, which will be performed in future work(for example using the EDEN simulations; \citep[][]{2025ApJ...986..147W}).

To account for baryonic effects on the disruption of satellite galaxies in our neural network framework, we use the UM-SAGA catalog \citep[][]{Wang2024}, which is a modification of the \textsc{UniverseMachine} (UM) galaxy-halo connection framework. The \textsc{UniverseMachine} \citep{2019MNRAS.488.3143B} models the relationship between galaxies and their dark matter halos based on observable data. Specifically, it learns the relationships between halo mass, assembly history, and galaxy star formation rates (SFRs) that are necessary to match observed data, including galaxy number densities, correlation functions, star formation rates, quenched fractions, luminosity functions, among others. Satellite lifetimes are truncated or extended based on the fraction of their maximum circular velocity that they have lost since infall; satellites that have been lost in the underlying dark matter simulation are assumed to continue orbiting as point particles in their host dark matter halos with the mass loss prescription of \cite{Jiang16} until they reach the requisite circular velocity threshold.  This satellite disruption threshold is adjusted via Markov Chain Monte Carlo until it (along with other star formation parameters) produces mock universes consistent with observations.  Specifically, the primary constraining power for the lifetimes of satellite galaxies comes from a combination of 1) galaxy number densities and large-scale ($>1$ Mpc) clustering (which set the overall stellar mass--halo mass relation), and 2) small-scale ($<1$ Mpc) auto- and cross-correlation functions, which determine the present-day fraction of satellites at a given stellar mass.  

Although this method can constrain satellite lifetimes in principle, the original \textsc{UniverseMachine} only considered clustering data for $M_\ast > 10^{10.3}\Msun$, i.e., substantially larger masses than most MW satellites. The UM-SAGA model incorporates all the data used to constrain the original \textsc{UniverseMachine}, along with data from the SAGA Survey  (Satellites Around Galactic Analogs; \citealt{Geha2017,Mao2021}) and SDSS (Sloan Digital Sky Survey; \citealt{York2000}) isolated field galaxies, vastly improving the data constraints for galaxies with $M_\ast < 10^9\Msun$. These constraints include the stellar mass function of SAGA satellite galaxies, quenched fractions (i.e., fraction of galaxies that have stopped forming stars) for both SAGA satellites and SDSS isolated galaxies as a function of stellar mass, the relationship between satellite quenched fractions and the distance to the host galaxy, and specific star formation rates (SFR) for star-forming satellites.   The additional data most relevant for satellite lifetimes are the numbers of spectroscopically confirmed neighbors around MW-like galaxies in the SAGA sample, which are complete down to a stellar mass of $\sim 10^{7.5}\Msun$ \citep{2024ApJ...976..117M}.  This allows UM-SAGA to better fit the distribution of neighbors (both in absolute number and in radial profile) around Milky Way-like systems.


Both the \textsc{UniverseMachine} and UM-SAGA produce catalogs of galaxies (both satellite and central) when applied to a dark matter simulation, along with inferred stellar masses and star formation rates.  In this paper, we use primarily the halo catalogs output by both methods(\textsc{UniverseMachine} and UM-SAGA), as we do not directly compare to observed galaxies' stellar masses or star formation rates.  As discussed in Section \ref{obs:errors}, we use stellar masses only indirectly, and add significant noise to limit any information leakage from the stellar mass--halo mass relationship to our mass estimates.  We show comparisons for both UM-SAGA and \textsc{UniverseMachine} catalogs to estimate the systematic uncertainties for baryonic effects on satellite disruption, but we take results from the UM-SAGA catalog to reflect our best mass estimate.


\begin{table*}
\centering
\begin{tabular}{l c | c p{55pt} | c p{55pt}}
\hline
Simulation/Model & Particle Resolution & Hosts with $\ge 10$ neighbors & Median $M_\mathrm{vir}$ in Training Sample & Hosts with $\ge 25$ neighbors & Median $M_\mathrm{vir}$ in Training Sample\\
\hline
\hspace{-1.5ex}$^*$ESMDPL/UM-SAGA & $3.25 \times 10^{5}\Msun/h$ & 83,270 &  11.14 & 29,733 & 11.82\\
ESMDPL/UM & $3.25 \times 10^{5}\Msun/h$ & 94,234 & 11.07 & 35,293 & 11.70\\
VSMDPL/UM-SAGA &  $6.2 \times 10^{6}\Msun/h$ & 541,594 & 11.91 & 140,392 & 13.23 \\
VSMDPL/UM &  $6.2 \times 10^{6}\Msun/h$ & 679,802 & 11.68 & 165,862 & 13.07\\
\hline
\end{tabular}
\caption{\normalfont Sample sizes from the simulations/models used.  ESMDPL=Extremely Small MultiDark Planck; VSMDPL=Very Small MultiDark Planck \citep{Klypin16,RP16}; UM=\textsc{UniverseMachine} \citep{2019MNRAS.488.3143B}; UM-SAGA=\textsc{UniverseMachine}-SAGA \citep{Wang2024}.  $^*$Results from the ESMDPL/UM-SAGA combination are expected to be most accurate, due to a combination of higher resolution and better treatment of satellites.  The UM and UM-SAGA models---for the purposes of this paper---will give different results primarily due to their different treatment of satellite lifetimes.}
\label{tab:simulation_props}
\end{table*}

\subsection{Halo Selection and Input Features}

We train deep neural networks to estimate host halo masses using halos with peak masses greater than $10{^8}\Msun$ from the UM and UM-SAGA galaxy catalogs, from both the ESMDPL and VSMDPL simulations. Halos with $M_\mathrm{vir}>10^{8}\Msun$ are the only ones expected to host galaxies with measurable proper motions, as star formation is strongly suppressed in lower-mass halos \citep[e.g.,][]{OShea15}. Unlike previous studies, no prior assumptions about the host halo mass of the Milky Way are made, and host selection is based solely on observable properties.

Many past approaches have also assumed that nearby galaxies were satellites of the Milky Way, strongly influencing host halo mass estimates. As in \citet[][]{2024OJAp....7E..74H}, we remove that assumption and instead use the orbital properties (radial distance $R$, proper motion $\mu$, and radial velocity $V_\mathrm{los}$) of nearby halos (whether satellites or not) as input features. Neighboring halos are selected out to 250 kpc, corresponding to the distance where proper motions of Milky Way satellite candidates can be robustly measured.  To reduce contamination from high-velocity satellites of neighboring massive halos, we restricted the training sample to include only host halos with distances of $>600$kpc to the nearest larger halo.

We train our neural networks using the orbital properties of 10 and 25 neighboring halos around the host halo.  Requiring a minimum number of neighbors imposes an additional selection effect, in that only halos that are massive enough or in dense enough environments to have $\ge 10$ or $\ge25$ neighbors with $M_h\ge10^{8}\Msun$ within $R=250$kpc can be selected. Host halo sample sizes and median masses are shown in Table \ref{tab:simulation_props}.  No prior constraints are applied to the mass of the central halo, allowing halo masses to range broadly from approximately $\sim 10^8 - 10^{15}\Msun$.  For halos with more neighbors than the required minimum, we use the 10 or 25 neighboring halos with highest stellar masses, as these would be those most likely to be selected observationally.  In practice, this is roughly equivalent to selecting the surrounding halos with highest $v_\mathrm{max}$ as the rank order in stellar mass strongly tracks the rank order in $v_\mathrm{max}$ \citet[][]{2023ApJ...948..104P}. We adopt 25 neighbors in this work (instead of 30 as used in \citet[][]{2024OJAp....7E..74H}) because the ESMDPL simulation has higher resolution but a smaller volume,  which results in fewer host halos that satisfy our detectability cut with $\geq30$ neighbors within 250 kpc. Requiring 30 neighbors would therefore reduce the eligible host sample and increase the prediction error.

In \cite{2024OJAp....7E..74H}, we found that using neighboring halo orbits alone leads to unreliable mass estimates for low-mass halos (particularly $M_\mathrm{vir}<10^{11}\Msun$). Additional dynamical information, such as the maximum circular velocity ($v_\mathrm{max}$) of the most massive satellite, helps to improve accuracy, especially in distinguishing between low-mass and high-mass host halos. The distance to the nearest larger halo ($D_\mathrm{larger}$) and the distance to the nearest larger halo with $M_\mathrm{vir}\ge 10^{14}\Msun$ ($D_\mathrm{14}$) are also included as input features, as nearby halos impact the surrounding potential well depth and hence the angular momentum distributions of the satellites.
Further explanation of these choices for inputs, as well as the relative importance of each, are discussed in our previous paper \citep[][]{2024OJAp....7E..74H}. While earlier analyses suggested that the angular momentum distribution of neighboring galaxies provides strong constraints on Milky Way–mass halos (see also \citealt{Patel17}), more recent work (e.g., \citealt[][]{2024OJAp....7E..50K}) has argued that angular momentum alone is not necessarily the best indicator of halo mass. In practice, especially when observational uncertainties are included, quantities like satellite radius, tangential velocity ($v_\mathrm{tan}$), and radial velocity ($v_\mathrm{rad}$) contribute roughly equally to constraining the host halo mass (see Appendix~\ref{a:mw_constraints}).

\subsection{Network Training}
\label{s:training}

Neural networks are composed of interconnected nodes that are arranged in layers, allowing them to learn complex patterns and relationships from data. In our study, we employed a deep, fully connected neural network (NN) to perform regression for estimating halo mass using galaxy observables. 

\begin{figure}
  \centering
  \includegraphics[width=1\columnwidth]{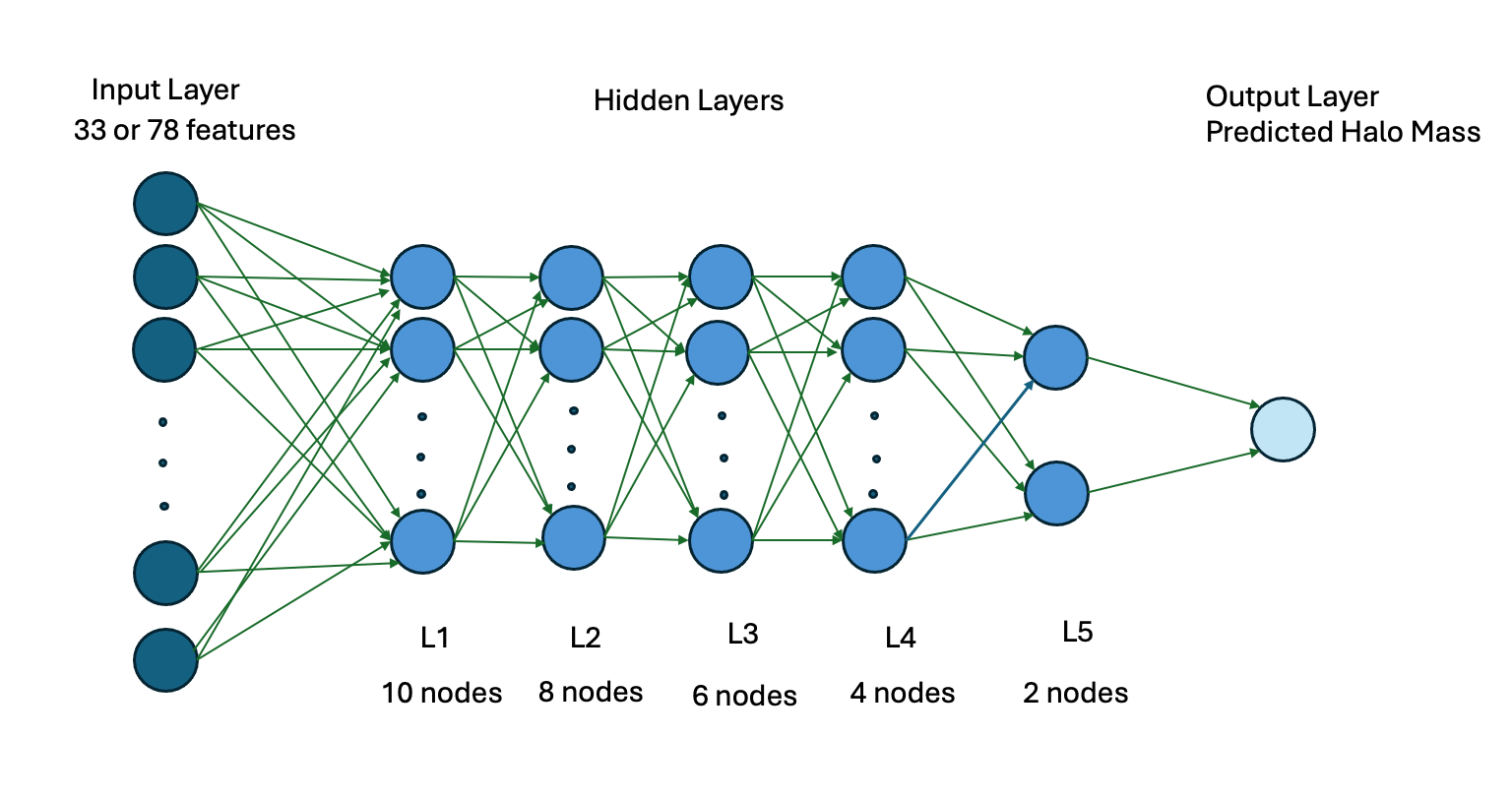}
  \caption{The neural network geometry we use to predict halo masses. Input features include neighboring halos' radial distance $R$, proper motion $\mu$, and radial velocity $V_\mathrm{los}$, as well as the maximum circular velocity of the most massive satellite ($v_\mathrm{max,sat}$), the distance to the nearest larger halo ($D_\mathrm{larger}$), and the distance to the nearest halo with $\log_{10}(M_\mathrm{vir}/\Msun)>14$ ($D_\mathrm{14}$). For all networks (regardless of the number of inputs), there are 5 hidden layers, gradually decreasing from 10 nodes to 2 nodes, with one output layer corresponding to the predicted halo mass.  (Figure graphic reproduced from \citealt{2024OJAp....7E..74H}.)}
  \label{fig:Neural}
\end{figure}

We reuse the network architecture and hyperparameters from our previous paper \citep[][]{2024OJAp....7E..74H}, where we explored different choices and found that they did not significantly influence the network performance.  The choices adopted in the previous work included: 
\begin{enumerate}
\item Input Size: The input layer includes three features related to the orbital properties (radial distance $R$, proper motion $\mu$, and radial velocity $V_\mathrm{los}$) of each neighboring halo, plus an additional three features regarding the target halo's environment: 1) $v_\mathrm{max}$ of the most massive satellite, 2) distance to the nearest larger halo ($D_\mathrm{larger}$), and distance to the nearest $10^{14}\Msun$ halo ($D_{14}$).  Here, we define $D_{\rm larger}$ as the distance to the nearest halo with $M_{\rm vir}$ greater than that of the target halo, and $D_{14}$ as the distance to the nearest halo with $M_{\rm vir}\geq10^{14}\,M_\odot$. Therefore, the network has 33 input features 
for 10 neighbors and 78 input features 
for 25 neighbors.
\item Layer Architecture: The architecture (Fig.\ \ref{fig:Neural}) consists of five fully-connected hidden layers. Each hidden layer applies a nonlinear transformation (in our case, a rectified linear unit; ReLU) to the sum of connected outputs from the previous layer. In a fully connected layer, each neuron receives inputs from all neurons in the preceding layer. Our initial hidden layer contains ten neurons. As we progress through the network, the number of neurons in subsequent layers decreases (8, 6, 4, 2), which reduces the information passed through each layer to that most relevant for host halo mass.
\item Output Layer: The output layer consists of a single linear layer to predict the single output value of the central halo mass.
\item Loss Function: Mean Squared Error (MSE).
\item Optimizer: Adam, which is an adaptive learning rate optimization algorithm. It adaptively determines the step size along the gradient of the loss function for each parameter update, often achieving faster convergence than traditional stochastic gradient descent (SGD).
\item Learning Rate: We set to 0.001, governing the initial optimization step size in Adam.
\item Batch Size: A batch size of 64 was employed, i.e., the number of halos used per weight update during network training.
\end{enumerate}


We used halos for simulation snapshots from $z=0$ to $z=0.25$ from the VSMDPL and ESMDPL simulations to add a greater variety of neighboring halo orbital configurations. Importing training data from earlier snapshots does not introduce a discernible bias in the median predicted masses for halos at $z=0$, implying consistency in the orbital configuration distribution over this redshift interval \citep{2024OJAp....7E..74H}. 

To limit information leakage between the training and test sets across different snapshots, we split halos based on spatial position: halos with X-coordinate less than $40$ Mpc/$h$ were assigned to the training set and those greater than $40$ Mpc/$h$ to the test set for ESMDPL (box size $64$ Mpc/$h$), and we used a threshold of $101$ Mpc/$h$ for VSMDPL (box size $160$ Mpc/$h$). This spatial split helps ensure that halos from similar regions do not appear in both sets, preserving the independence of the test sample.
 
 For pre-processing, we sorted neighboring halos by descending stellar mass as assigned by UM-SAGA or UM, applied observational errors to all input features as per Section \ref{obs:errors},  took the logarithm of all mock-observed input features, normalized by subtracting the mean across all neighbors, and scaled to unit variance. Subsequently, two 5--layer fully connected neural networks were trained on the 10-- and 25--neighbor mock observational input feature vectors to estimate the masses of the corresponding central halos for each simulation/model pair in Table \ref{tab:simulation_props}.  We repeated this training process 100 times for each simulation/model pair, and averaged the predictions, both to reduce variance and to increase reproducibility of our findings.  To visualize prediction errors, we computed the root mean square error (RMSE) in binned mass intervals of our network ensembles applied to halos in the test set---i.e., halos that had never been seen during training.   


 

\section{Observational Data and Error Model}
\label{s:data}

\subsection{Proper motions and radial velocities for Milky Way satellites}

Observational data includes the right ascension, declination,  distance modulus, and radial velocity of classical dwarfs and faint dwarfs of the Milky Way from \citet[][]{2020AJ....160..124M}, and both components of the proper motion from \citet[][]{2020RNAAS...4..229M}.  For the error model in Section \ref{obs:errors}, we adopted estimated errors in proper motion, radial velocity, and distance modulus from the above sources. We excluded streams from being considered as satellites, since disrupted sources are removed in the UM and UM-SAGA.

The properties of the LMC and SMC were obtained from separate works, including \citet[][]{2023A&A...669A..91J} for the proper motions of the LMC and the radial velocity of the SMC, \citet[][]{2013ApJ...764..161K} for the distance modulus of the LMC, \citet[][]{2019MNRAS.483..392D} for the radial velocity of the LMC, and \citet[][]{2018ApJ...864...55Z}, for the proper motions and the distance modulus of the SMC. 
We converted all position and velocity data to galactocentric coordinates using \texttt{astropy}, assuming the default solar position and velocity data: \citet[][]{2018RNAAS...2..210D} and  \citet[][]{2004ApJ...616..872R} for the solar velocity, \citet[][]{2018A&A...615L..15G} for the distance, and \citet[][]{2019MNRAS.482.1417B} for the solar position in the Milky Way.  All input satellite data is summarized in Appendix \ref{a:satellites}.

\subsection{Environment of the Milky Way}

\label{s:environment}

We assume a distance of 761 kpc between M31 and the MW as observed in \citet[][]{2021ApJ...920...84L}.  The assumed observed distance to the Virgo cluster is 17 Mpc, from \citet[][]{2008MNRAS.389.1539F}. Finally, the observed maximum circular velocity of the LMC is assumed to be 91.7 km/s, from  \citet{2013ApJ...764..161K}.   We use observational errors from these sources in the error model introduced in Section \ref{obs:errors}.

\subsection{Satellite Detectability}
\begin{figure}
  \centering
  \includegraphics[width=\columnwidth]{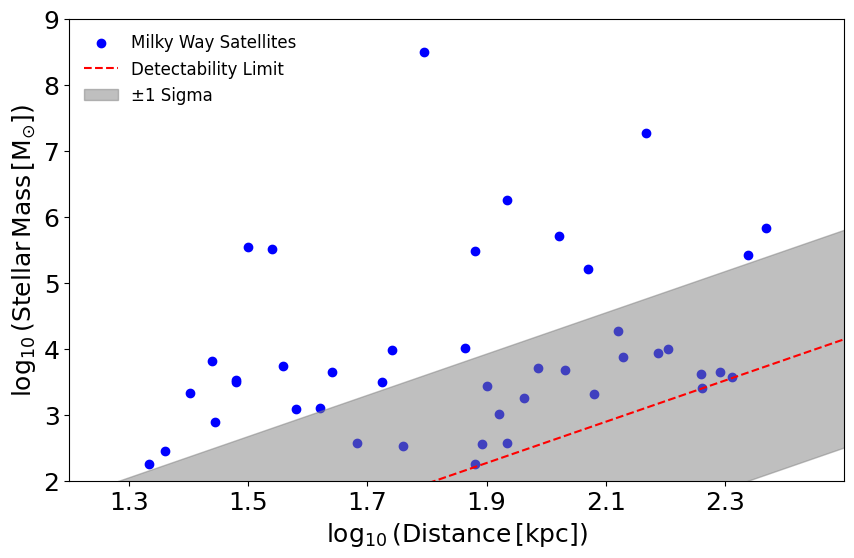}
  \caption{Distribution of satellite stellar masses as a function of distance. The detectability threshold (red dashed line) represents the mass cut applied to mimic observational limitations. The shaded region corresponds to the $\pm 1\sigma$ uncertainty in the selection function.}
  \label{fig:detectibility}
\end{figure}

Observational surveys of the Milky Way satellites exhibit a selection effect in which faint satellites are more difficult to detect at larger distances. This bias arises because low-luminosity satellites have lower surface brightnesses, making them harder to identify in surveys with finite depth. Figure ~\ref{fig:detectibility} illustrates this selection effect, where the distribution of Milky Way satellites is notably missing sources at low luminosities and large distances.  To mimic this effect in our simulated catalogs, we applied a detectability threshold based on a stellar mass cut:
\begin{equation} 
\log_{10} \left( M_* / M_{\odot} \right) = -3.7 + R(1) + \log_{10} \left( d / \text{kpc} \right) \times 3.125,  \label{e:cut}
\end{equation}
where $M_*$ is the stellar mass, $d$ is the satellite distance, and $R(1)$ represents a Gaussian random variable with a standard deviation of one dex. The intercept of $-3.7$ and the slope of $3.125$ were empirically chosen to approximate the observed absences of faint satellites at large distances in the Milky Way, i.e., they were calibrated against current MW satellite data rather than fitted freely. The scatter parameter corresponds to the standard deviation of the Gaussian random variable and is applied once per host halo.  This results in large scatter in the detectability threshold that is applied to all neighboring halos.  We adopt this method to limit information leaking from the UM or UM-SAGA stellar mass--halo mass relation into the selection of host halos for training---otherwise, lower-mass host halos would be uniformly excluded, as they would not be surrounded by enough bright neighbors to be included in our training samples. The red threshold line in Fig.\ \ref{fig:detectibility} represents the detectability threshold, and the shaded region corresponds to the $\pm 1\sigma$ range in the selection cut in Eq.\ \ref{e:cut}. 

\subsection{Simulating Observational Errors}

\label{obs:errors}
Here, we describe how we model the distribution of observational errors for Milky Way satellites based on their distances and luminosities.  As described in Section \ref{s:training}, we apply this error model to simulated halos, to create mock observations, use the mock observations of simulated halos to train the neural networks, and then apply the neural networks to the observations to estimate the mass of the Milky Way.  Of note, we do not resample observed quantities within their uncertainties, as this would artificially increase the dispersion of orbital properties and lead to a mass bias.

We use linear regressions to model the typical observational errors for three observational quantities: distance, proper motion, and radial velocity. For each quantity, we fit the errors with the following equation:
\begin{equation}
    \log_{10}(\text{E}) = a\cdot\log_{10}(\text{D}) + b\cdot\log_{10}(M_*) + c + R(\sigma),
\end{equation}
where $E$ represents the error for the respective quantity (relative distance, proper motion, or radial velocity), and $a$ (slope with log distance), $b$ (slope with log stellar mass), and $c$ (intercept) are coefficients determined using least-squares optimization, $R(\sigma)$ captures residual scatter from the relation assuming a log-normal standard deviation $\sigma$, $D$ is the distance in kpc, and $M_*$ is the stellar mass in $\Msun$.  For most satellites, only $v$-band absolute magnitudes were available. In the absence of color information, stellar mass estimates become more uncertain because $M_\ast$ is correlated with color in surveys such as SAGA. We account for this additional source of scatter by folding it into the magnitude-dependent error term $b$, which effectively absorbs the increased uncertainty from using single-band photometry. To estimate stellar masses for the remaining cases, we convert absolute magnitudes using an assumed mass-to-light ratio of 1.

\begin{figure*}
    \begin{minipage}[t]{0.48\textwidth}
        \centering
        \includegraphics[width=\linewidth,valign=c]{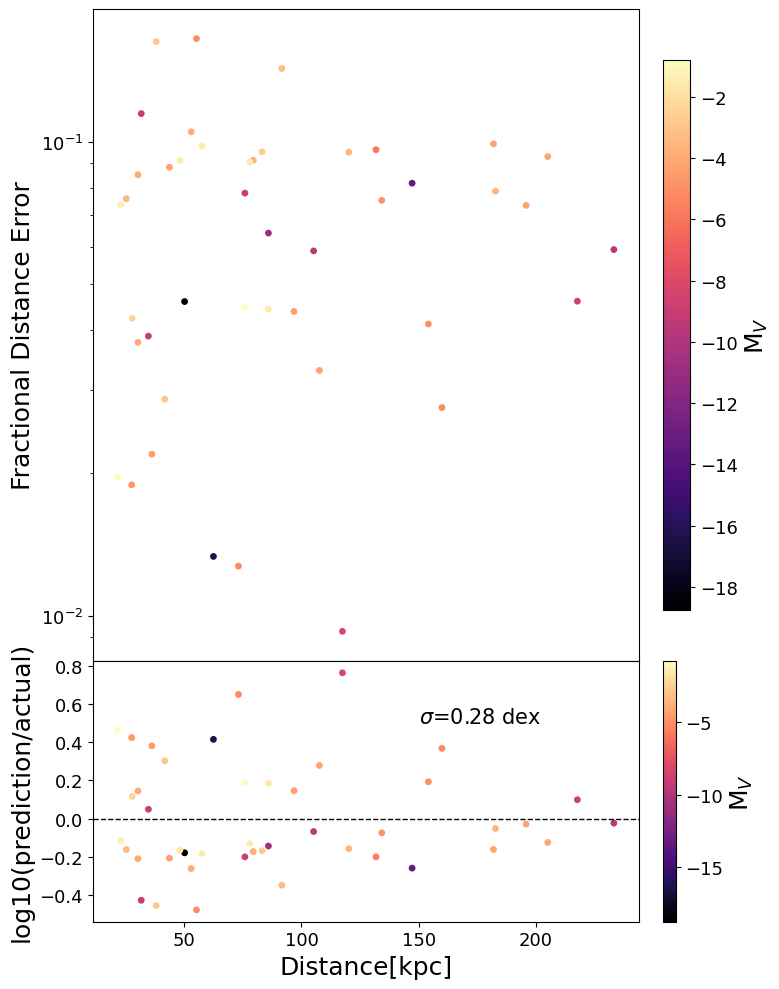}
        \caption{Fractional distance errors for galaxies nearby the Milky Way as a function of their heliocentric distance and their luminosity. Satellite luminosity is represented by the color bar. The fractional distance error does not exhibit a significant trend with luminosity. However, there is a selection bias, as galaxies with lower luminosity become harder to detect at larger distances.}
        
        \label{fig:distance_error}
    \end{minipage}\qquad
    \begin{minipage}[t]{0.48\textwidth}
        \centering
        \includegraphics[width=\linewidth,valign=c]{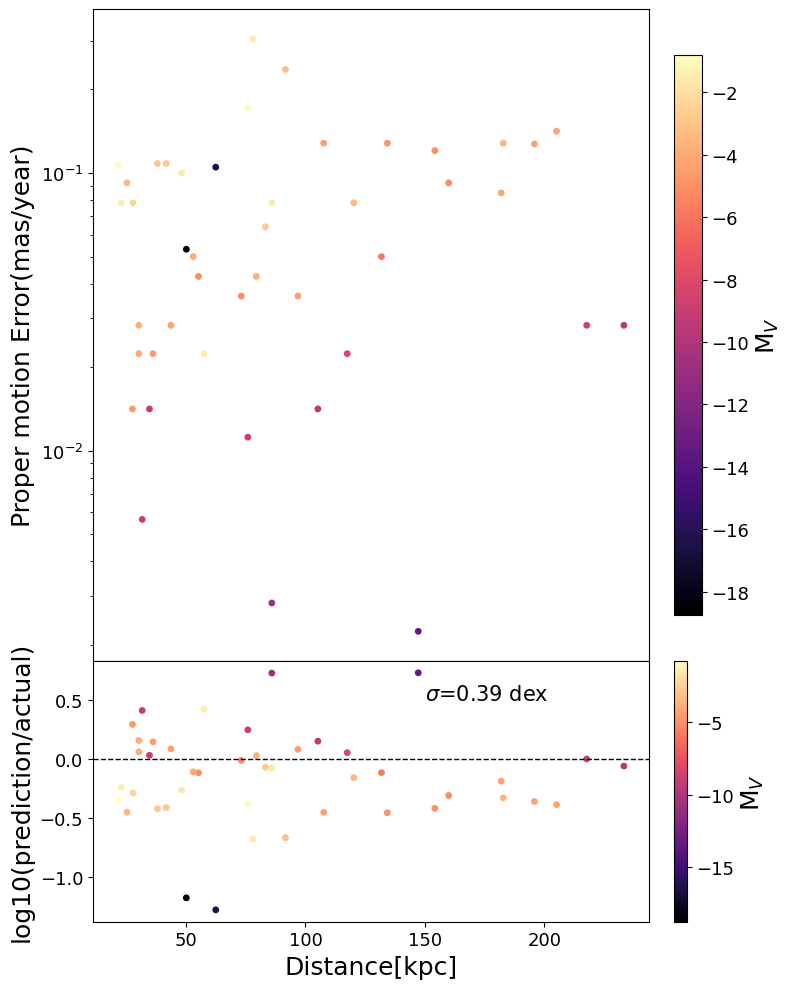}
        \caption{Proper motion errors for galaxies nearby the Milky Way as a function of their heliocentric distance and luminosity. There is a trend indicating that greater brightness is associated with lower error rates. However, the Large Magellanic Cloud (LMC) and Small Magellanic Cloud (SMC) likely represent exceptions, maintaining relatively high errors (see dark blue points in bottom plot) due to their more complex structures and dynamics compared to other galaxies.}
        \label{fig:proper_mption_error}
        \end{minipage}
\end{figure*}

\begin{figure*}
\begin{minipage}[t]{0.48\textwidth}
\includegraphics[width=\columnwidth]{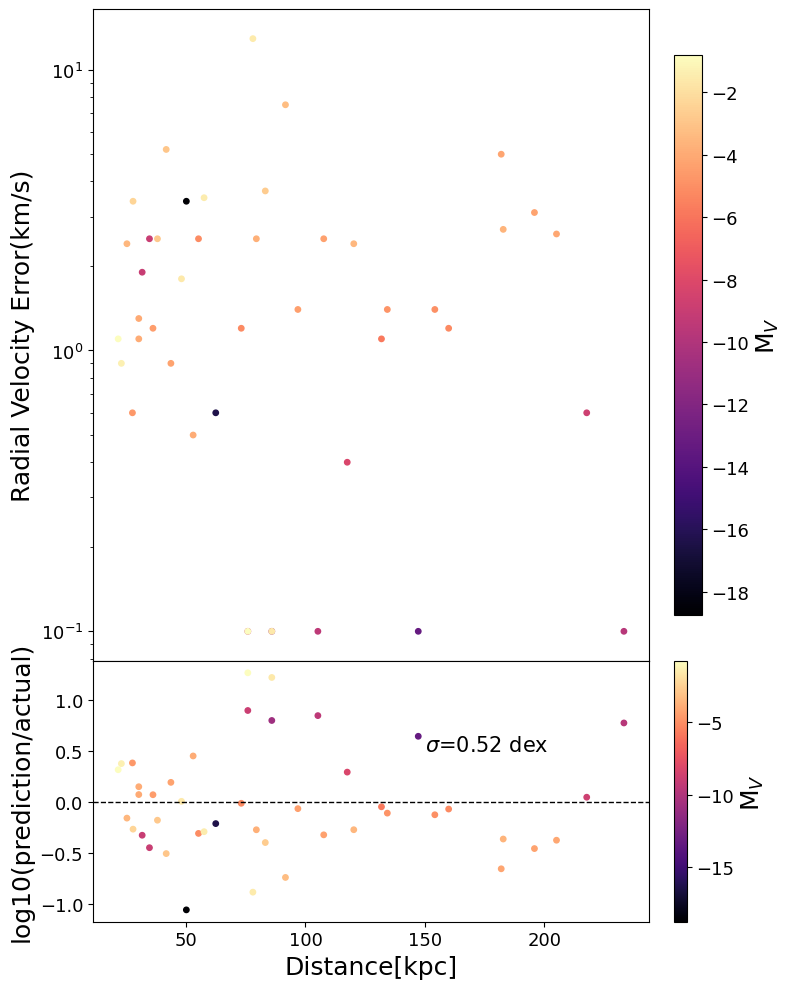}

  \caption{The plot shows the radial velocity error for Milky Way satellite halos as a function of their luminosity and the distance to their host halo. Also there isn't a strong trend with luminosity in terms of the typical errors.}
  
  \label{fig:radial_velocity_error}
\end{minipage}\qquad
    \begin{minipage}[t]{0.48\textwidth}  \centering
  \includegraphics[width=1\columnwidth]{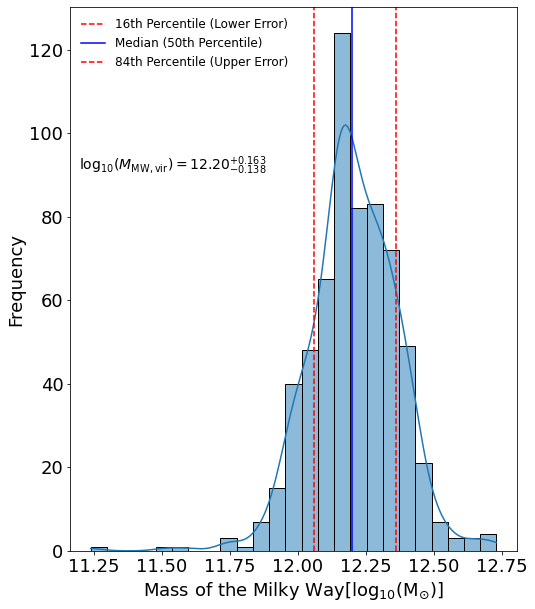}
  \caption{Posterior distribution of virial masses for the Milky Way based on the ESMDPL/UM-SAGA catalog. The blue line indicates the median mass and the two dashed lines show the 16$^\mathrm{th}$ and $84^\mathrm{th}$ percentiles.}
  \label{fig:histogram}
  \end{minipage}

  \begin{minipage}{0.5\textwidth}
    \centering
    \includegraphics[width=\columnwidth]{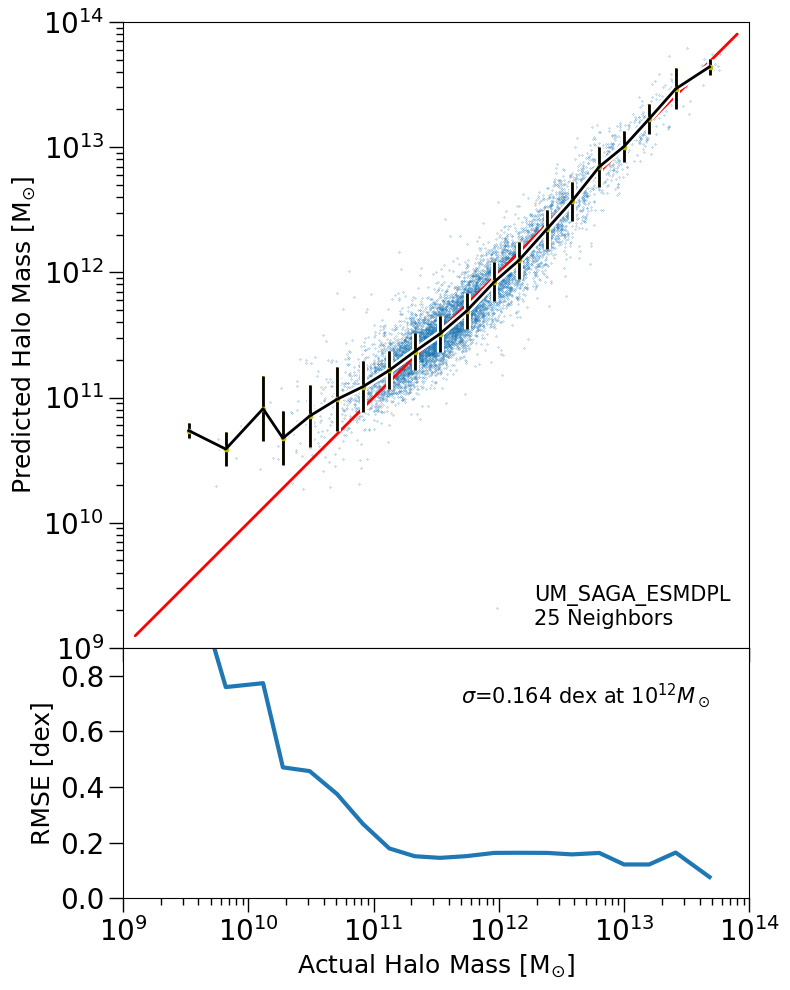}
  \end{minipage}%
  \begin{minipage}{0.5\textwidth}
    \centering
    \includegraphics[width=\columnwidth]{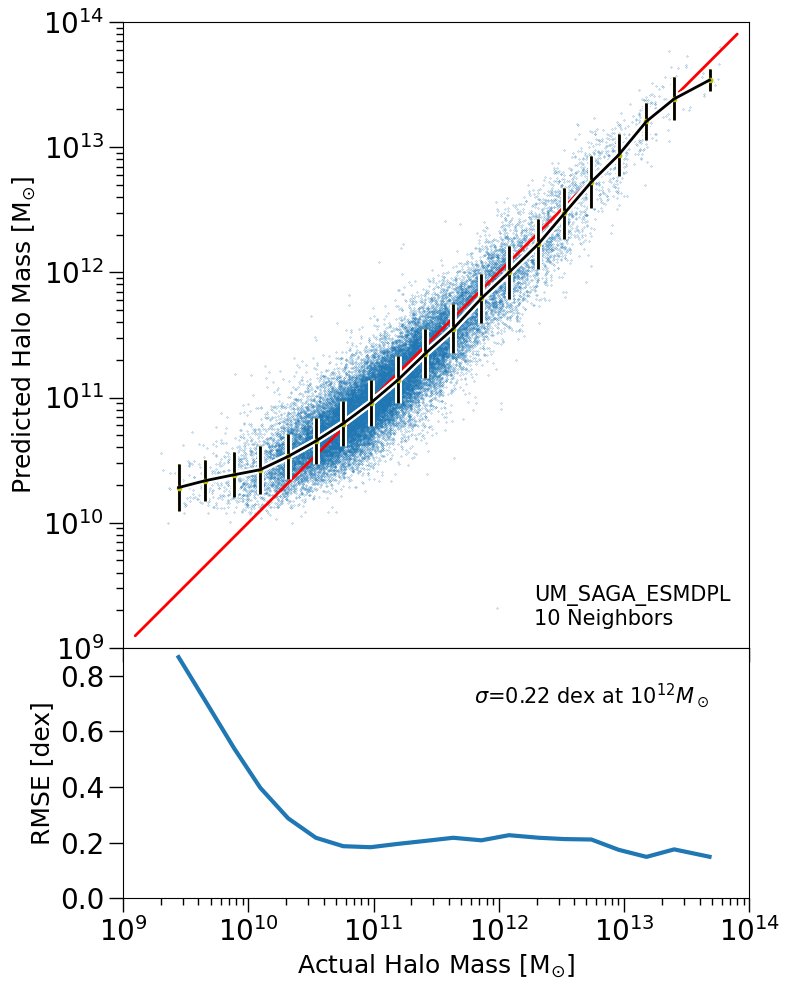}
  \end{minipage}
  \caption{The comparison between predicted and actual halo masses by neural networks using the UM\_SAGA\_ESMDPL catalog. The left-hand figure shows results for halos with at least 25 surrounding galaxies, with an error of approximately 0.16 dex, while the right-hand figure shows results for halos with at least 10 neighboring galaxies, with an error of 0.22 dex. The bottom panels display the root mean square error (RMSE) as a function of actual halo mass. Error bars indicate the standard deviations of the predicted mass in accordance with the actual mass. The black line signifies the median of predicted masses across different bins of actual mass, while the red line serves as a reference where the predicted mass matches the actual mass.}
  \label{fig:result_saga}
  \vspace{0.5cm} 

\end{figure*}

We show the overall dependence on luminosity and distance of each orbit observable in Figs.~\ref{fig:distance_error}, ~\ref{fig:proper_mption_error}, and ~\ref{fig:radial_velocity_error}.  We do not show fits directly, as they are two-dimensional functions, but instead show residuals to the median predicted error in the lower panel. The fitting parameters for the different errors are presented in Table ~\ref{tab:fitting_parameters}.

The observed trends in these figures align with expectations based on observational constraints. The fractional distance error (Fig.~\ref{fig:distance_error}) does not show a strong dependence on luminosity, which is expected since distance uncertainties in surveys like \textit{Gaia} are primarily influenced by parallax precision and photometric depth rather than brightness.  However, \textit{Gaia}'s precision is not sufficient for measuring the bulk proper motions of low-mass systems beyond $\sim$400 kpc, and its magnitude limit of $G \sim 20.7$ mag means that some more distant satellites will remain permanently out of reach. Proper motion errors (Fig.~\ref{fig:proper_mption_error}) decrease somewhat with increasing brightness, consistent with \textit{Gaia}’s astrometric performance, where brighter sources have more precise proper motion measurements. Notably, the Large Magellanic Cloud (LMC) and Small Magellanic Cloud (SMC) exhibit higher-than-expected proper motion errors, likely due to their complex internal kinematics and extended structures, which introduce additional measurement uncertainties. In contrast, radial velocity errors (Fig.~\ref{fig:radial_velocity_error}) show no strong dependence on luminosity or distance, which aligns with expectations since radial velocities are measured spectroscopically, providing high precision even for faint sources as long as the spectral signal-to-noise ratio is sufficient. The scatter in radial velocity errors likely reflects variations in spectrograph resolution and target selection rather than an intrinsic correlation with luminosity or distance. In addition, spectral signal to noise ratio (SNR) is correlated with star formation rate: star-forming satellites with strong emission lines yield more precise velocity measurements, whereas quenched systems are harder to fit with spectral templates. A secondary effect arises from satellite morphology, as fiber placement on a quenched bulge can reduce the SNR even for galaxies with star-forming outskirts. These results suggest that our error modeling provides a reasonable match to physical expectations.

To account for observational uncertainties in key environmental factors, we introduce Gaussian noise into the measured values of the distance to the nearest larger halo (DNL), the distance to the nearest $10^{14} M_{\odot}$ halo (D14), and the maximum circular velocity of the most massive satellite (Vmax). The nearest larger halo to the Milky Way is Andromeda, with an observed distance of 761 kpc. To model uncertainties in this measurement, we apply a Gaussian perturbation with a standard deviation of 0.007. Similarly, the Virgo cluster, the closest halo with $M_{\text{vir}} \geq 10^{14} M_{\odot}$, is located at an observed distance of 17 Mpc. The uncertainty in D14 is modeled as a log-normal distribution with a standard deviation of 

\[
\sigma_{\log} = \log_{10} \left(1 + \frac{0.25}{17} \right),
\]
capturing fractional errors in distance estimates. For the maximum circular velocity of the Large Magellanic Cloud (LMC), which is observed to be 91.7 km/s, we introduce a Gaussian error with a standard deviation of 5.95 km/s to reflect measurement uncertainties. These uncertainties are applied as random noise to the simulated dataset, ensuring that our machine learning models are trained on data that realistically reflect observational limitations.

\begin{table*}
\centering
\begin{tabular}{|c|c|c|c|c|}
\hline
Model & Intercept (c) & Slope with Distance (a) & Slope with Stellar Mass (b) & Scatter ($\sigma$)\\ \hline
Distance & -1.34 & 0.15 & -0.05 & 0.28\\ \hline
Proper Motion & -1.29 & 0.35 & -0.16 & 0.39 \\ \hline
Radial Velocity & 0.84 & -0.17 & -0.11 & 0.52 \\ \hline
\end{tabular}
\caption{Fitting parameters for the different observational error models. The equation used for the models is given by:
\[
\text{Intercept} + (\text{Slope with Distance}) \cdot \log\left(\frac{\text{Distance}}{\text{kpc}}\right) + (\text{Slope with Stellar Mass}) \cdot \log\left(\frac{\text{Stellar Mass}}{\text{M}_{\odot}}\right)
\]
}
\label{tab:fitting_parameters}
\end{table*}

 

\section{The total mass of the Milky Way}
\label{s:results}

In this section, we present the results of applying our trained neural networks to simulated and observed data. We first evaluate the network performance on the ESMDPL/UM-SAGA catalog, analyzing the accuracy and root-mean-square errors (RMSE) of the predicted halo masses. Then, we report the inferred Milky Way mass using different simulation and satellite catalog combinations and discuss the systematic uncertainties introduced by these choices.

Figure \ref{fig:result_saga} shows comparisons between predicted and actual mass for the ESMDPL/UM-SAGA catalog, and similar figures for the remaining catalogs are shown in Appendix \ref{a:catalog_pairs}.
 The median values of halo masses predicted by the neural networks, when grouped by actual halo mass, align well with the true halo masses, typically showing median discrepancies of no more than 0.05 dex at halo masses of $10^{12}\Msun$. The networks' root-mean-square errors (RMSE) show a characteristic drop and plateau as a function of actual halo mass. The RMSE shapes are largely impacted by whether the neighboring galaxies within a 250 kpc radius are satellites.  For the ESMDPL/UM catalog, the RMSE for 25 satellites reaches its minimum value of approximately 0.160 dex at $10^{12}$ M$_\odot$, while for 10 satellites, the minimum RMSE was 0.199 dex. This demonstrates that larger satellite samples provide more reliable mass estimates, with the reduction close to that expected for Poisson noise ($\sqrt{2.5}$).  Observational errors contribute significantly to these uncertainties; in our previous research \citet[][]{2024OJAp....7E..74H}, we found that perfect measurements would result in uncertainties of only 0.118 dex for 25 satellites and 0.165 dex for 10 satellites. 
 
 We found that the training process introduced variance of approximately 0.05 dex between single training runs with different initial random seeds, so we would expect the training variance in our 100-run average to be 0.005 dex. It is worth noting that the RMSE of $0.164$ dex for the ESMDPL/UM-SAGA at $10^{12}\Msun$ is slightly larger than 0.153 dex reported in Fig.\ \ref{fig:histogram} due to a small offset (0.05 dex) between the average predicted and true mass. Additionally, the 0.153 dex corresponds to the 68\% confidence interval, whereas the RMSE accounts for the influence of high- and low-mass outliers.

In this work, we applied observational errors and selection effects to the simulated halo catalogs and trained neural networks to infer total halo masses from observables.  We then applied these networks to actual observations for both dwarf galaxies around the Milky Way (MW) and the environmental variables in Section \ref{s:environment} to estimate the total mass (dark+baryonic) of the MW's halo.  As discussed in Section \ref{s:methods}, we trained our networks on two different simulations: 1) the high-resolution ESMDPL, and 2) the lower-resolution VSMDPL, using two different catalogs for each simulation: 1) the UM-SAGA catalog (trained to match the distribution of satellites around MW-like galaxies from the SAGA survey), and 2) the \textsc{UniverseMachine} (UM) Data Release 1 (only trained to match stellar mass functions, but no correlation functions for satellite-mass galaxies).  For each simulation/catalog pair, we trained two network ensembles, one using data from the brightest 10 neighboring galaxies, and the second using data from the brightest 25 neighboring galaxies, both within 250 kpc.  We view the ESMDPL/UM-SAGA/25 neighbor combination as the most accurate, but include results from the other simulation/catalog/neighbor choices to provide an estimate of systematic uncertainties. 

For the ESMDPL/UM-SAGA catalog, the inferred Milky Way halo masses for each simulation and satellite selection are summarized in Table~\ref{tab:mw_mass_results}. Our preferred result, based on the ESMDPL/UM-SAGA catalog with 25 satellites, is highlighted in bold. The results indicate that increasing the number of satellites improves the accuracy of the MW mass estimation, as reflected in the reduced scatter and smaller error bars in the plots. We also notice an increase in the MW mass estimate when we include more low-mass satellites (i.e., going from 10 to 25 satellites). We expect the more massive satellites (i.e., in the 10-satellite case) to have experienced stronger dynamical friction, which has more significantly altered their specific angular momenta compared to less massive satellites. In contrast, lower-mass satellites are less affected by dynamical friction and may therefore retain orbital properties that more faithfully trace the underlying potential well. Smaller uncertainties were achieved with the UM-SAGA catalogs, as these catalogs contain fewer satellites per MW halo on average, so the requirement for a minimum number of neighbors tended to select a narrower range of host halo masses.

\begin{table}[ht]
\centering
\caption{Inferred Milky Way virial mass $\log_{10}(M_h/M_\odot)$ from different simulations and satellite sets.}
\label{tab:mw_mass_results}
\begin{tabular}{lcc}
\hline
Catalog & 25 Satellites & 10 Satellites \\
\hline
ESMDPL/UM-SAGA & \textbf{12.20$^{+0.163}_{-0.138}$} & 12.07$^{+0.186}_{-0.175}$ \\
ESMDPL/UM      & 12.18$^{+0.141}_{-0.147}$ & 12.03$^{+0.151}_{-0.154}$ \\
VSMDPL/UM      & 11.98$^{+0.133}_{-0.137}$ & 11.70$^{+0.160}_{-0.169}$ \\
VSMDPL/UM-SAGA & 12.09$^{+0.140}_{-0.149}$ & 11.93$^{+0.167}_{-0.170}$ \\
\hline
\end{tabular}
\end{table}

\section{Discussion and Conclusions}

\label{s:discussion}

In this paper, we use a neural network to constrain the total halo mass of the Milky Way, primarily informed by the orbits of surrounding galaxies.  We placed tight constraints for the mass of the Milky Way ($\log_{10} M_\mathrm{vir}/\Msun=12.20^{+0.163}_{-0.138}$, $\log_{10} M_\mathrm{200c}/\Msun=12.14^{+0.163}_{-0.138}$) using recent \textit{Gaia} DR3 results.
 We compare our estimated Milky Way mass ($M_\mathrm{200c}$) with results from other studies \citep[i.e.,][]{1989ApJ...345..759Z,2018ApJ...865...72E,2018ApJ...857...78P,2018ApJ...862...52S,2019ApJ...873..118W,2019MNRAS.487L..72G,2019MNRAS.485.3514D,2019A&A...621A..56P,2019MNRAS.484.5453C,2020JCAP...05..033K,2020MNRAS.494.4291C,2020ApJ...894...10L,2020MNRAS.494.5178F,2020ApJ...888..114Z,2021MNRAS.501.5964D,2022ApJ...926..189N,2022ApJ...924..131S,2022ApJ...925....1S,2023ApJ...945....3S}
in Figure ~\ref{fig:comparison}, ordered by year of publication.

Our method provides several key advantages over previous approaches, addressing certain limitations while opening avenues for further advancements. Firstly, we demonstrate that it is not necessary to assume dynamical equilibrium or classify objects as satellites to place strong constraints on halo masses, at least for halos with a sufficient number of nearby satellites. Secondly, we do not assume independence of tracers, as is implict in many other techniques, and so naturally incorporate correlations between tracer orbits (e.g., satellites of satellites).  Lastly, we forward model how observational errors affect the underlying true values of the neighboring galaxy orbits, instead of the common method of drawing samples from the observed posterior distributions.  While the latter method is appropriate for a single object, it tends to inflate the scatter between different objects when applied to a population---leading, for example, to overestimating the velocity dispersion compared to the true underlying value.  While we do not claim the smallest error bars compared to past studies, we believe that our error bars represent an accurate assessment of what is achievable with present data.

Looking toward the future, we note that neural networks have the potential to combine arbitrary information to place constraints on halo mass.  For example, using galaxy stellar masses provides an alternative method to estimate host halo masses, with about 0.17 dex statistical uncertainties \citep{Bowden23}.  While we limited our input information to dynamics and distances due to large systematic uncertainties in estimating stellar masses, we note that there are many observables that correlate with halo mass beyond dynamical tracers that could be used in future networks.

Beyond halo mass estimation, we plan to develop neural networks that can predict additional properties, such as the halo’s  concentration, spin axis,and assembly history. These properties will provide  insight into the structure and evolution of our own Galaxy’s halo, particularly for satellite orbit modeling, as current models often assume a static mass and concentration history for the Milky Way.

\section*{Acknowledgments}

This research made use of data from the SAGA Survey (sagasurvey.org). The SAGA Survey is a spectroscopic survey with data obtained from the Anglo-Australian Telescope, the MMT Observatory, and the Hale Telescope at Palomar Observatory. The SAGA Survey made use of public imaging data from the Sloan Digital Sky Survey (SDSS), the DESI Legacy Imaging Surveys, and the Dark Energy Survey, and also public redshift catalogs from SDSS, GAMA, WiggleZ, 2dF, OzDES, 6dF, 2dFLenS, and LCRS. The SAGA Survey was supported by NSF collaborative grants AST-1517148 and AST-1517422 and by Heising–Simons Foundation grant 2019-1402.

EH thanks the LSSTC Data Science Fellowship Program, which is funded by LSSTC, NSF Cybertraining Grant \#1829740, the Brinson Foundation, and the Moore Foundation; her participation in the program has benefited this work.  PB was funded by a Packard Fellowship, Grant \#2019-69646. EP acknowledges financial support provided by NASA through the Hubble Fellowship grant \# HST-HF2-51540.001-A awarded by STScI. STScI is operated by the Association of Universities for Research in Astronomy, Incorporated, under NASA contract NAS5-26555. 

 This research is based upon High Performance Computing (HPC) resources supported by the University of Arizona TRIF, UITS, and Research, Innovation, and Impact (RII) and maintained by the UArizona Research Technologies department. The University of Arizona sits on the original homelands of Indigenous Peoples (including the Tohono O’odham and the Pascua Yaqui) who have stewarded the Land since time immemorial.

The authors gratefully acknowledge the Gauss Centre for Supercomputing e.V. (www.gauss-centre.eu) for providing computing time on the GCS Supercomputer SUPERMUC-NG at Leibniz Supercomputing Centre (www.lrz.de) for the VSMDPL and ESMDPL simulations. The CosmoSim data base (www.cosmosim.org) provides access to the VSMDPL and the ESMDPL simulations and the Rockstar data. The data base is a service by the Leibniz Institute for Astrophysics Potsdam (AIP).

GY  would like to thank Ministerio de Ciencia e Innovaci\'on (Spain) for 
financial support under the project grant PID2021-122603NB-C21.

EH would also like to thank Philip Pinto, Gurtina Besla, Haowen Zhang, Haley Bowden, and Himansh Rathore for their valuable feedback, guidance, and support throughout the course of this work.

\bibliographystyle{mnras}
\bibliography{mnras_template} 
\appendix

\section{Observed Satellite Data}

\label{a:satellites}
 
\begin{longtable}{lrrlllllll}
\toprule
        Galaxy &   RA [deg]  &   Dec [deg]  &                $(m - M)_{0}$ &               $V_h$ [km/s] &          $\mu_{\alpha}\cos\delta$ [mas/yr] &               $\mu_{\delta}$  [mas/yr] &             Distance [kpc] &       Abs.\ Magnitude & $M_*$ [$\Msun$] \\
\midrule
\endhead
\bottomrule
\endfoot
           LMC &      81.9100 &      $-69.8700$ &    $18.5^{+0.1}_{-0.1}$ &   $262.2^{+3.4}_{-3.4}$ &   $1.86^{+0.015}_{-0.015}$ &  $0.341^{+0.051}_{-0.051}$ &    $50.12^{+2.36}_{-2.26}$ &    $-18.75^{+0.00}_{0.00}$ &      $2.7 \times 10^8$ \\
           SMC &      15.2370 &      $-72.2730$ & $18.98^{+0.03}_{-0.03}$ &   $145.6^{+0.6}_{-0.6}$ &       $0.82^{+0.1}_{-0.1}$ &  $-1.21^{+0.032}_{-0.032}$ &    $62.52^{+0.87}_{-0.86}$ &  $-16.40^{+0.00}_{0.00}$ &   $3.1 \times 10^8$ \\
        Fornax &      39.9971 &      $-34.4492$ & $20.84^{+0.18}_{-0.18}$ &    $55.3^{+0.1}_{-0.1}$ &  $0.382^{+0.001}_{-0.001}$ & $-0.359^{+0.002}_{-0.002}$ & $147.23^{+12.72}_{-11.71}$ & $-13.34^{+0.14}_{-0.14}$ &   $1.9 \times 10^7$ \\
      Sculptor &      15.0392 &      $-33.7092$ & $19.67^{+0.14}_{-0.14}$ &   $111.4^{+0.1}_{-0.1}$ &  $0.099^{+0.002}_{-0.002}$ & $-0.160^{+0.002}_{-0.002}$ &    $85.90^{+5.72}_{-5.36}$ & $-10.82^{+0.14}_{-0.14}$ &   $1.8 \times 10^6$ \\
          Leo2 &     168.3700 &       22.1517 & $21.84^{+0.13}_{-0.13}$ &    $78.0^{+0.1}_{-0.1}$ &    $-0.14^{+0.02}_{-0.02}$ &    $-0.12^{+0.02}_{-0.02}$ & $233.35^{+14.40}_{-13.56}$ &  $-9.74^{+0.04}_{-0.04}$ &   $6.7 \times 10^5$ \\
        Carina &     100.4029 &      $-50.9661$ & $20.11^{+0.13}_{-0.13}$ &   $222.9^{+0.1}_{-0.1}$ &     $0.53^{+0.01}_{-0.01}$ &     $0.12^{+0.01}_{-0.01}$ &   $105.20^{+6.49}_{-6.11}$ &  $-9.45^{+0.05}_{-0.05}$ &   $5.2 \times 10^5$ \\
    UrsaMajor2 &     132.8750 &       63.1300 &    $17.5^{+0.3}_{-0.3}$ &  $-116.5^{+1.9}_{-1.9}$ & $-0.124^{+0.004}_{-0.004}$ &  $0.078^{+0.004}_{-0.004}$ &    $31.62^{+4.69}_{-4.08}$ &  $-9.03^{+0.05}_{-0.05}$ &   $3.5 \times 10^5$ \\
        Segue2 &      34.8167 &       20.1753 &    $17.7^{+0.1}_{-0.1}$ &   $-39.2^{+2.5}_{-2.5}$ &    $-0.41^{+0.01}_{-0.01}$ &     $0.04^{+0.01}_{-0.01}$ &    $34.67^{+1.63}_{-1.56}$ &  $-8.94^{+0.06}_{-0.06}$ &   $3.2 \times 10^5$ \\
         Draco &     260.0517 &       57.9153 &  $19.4^{+0.17}_{-0.17}$ &  $-291.0^{+0.1}_{-0.1}$ &  $0.042^{+0.005}_{-0.005}$ &    $-0.19^{+0.01}_{-0.01}$ &    $75.86^{+6.18}_{-5.71}$ &  $-8.88^{+0.05}_{-0.05}$ &   $3.0 \times 10^5$ \\
CanesVenatici1 &     202.0146 &       33.5558 &   $21.69^{+0.1}_{-0.1}$ &    $30.9^{+0.6}_{-0.6}$ &    $-0.11^{+0.02}_{-0.02}$ &    $-0.12^{+0.02}_{-0.02}$ &  $217.77^{+10.26}_{-9.80}$ &  $-8.73^{+0.06}_{-0.06}$ &   $2.7 \times 10^5$ \\
       Crater2 &     177.3100 &      $-18.4131$ & $20.35^{+0.02}_{-0.02}$ &    $87.5^{+0.4}_{-0.4}$ &    $-0.07^{+0.02}_{-0.02}$ &    $-0.11^{+0.01}_{-0.01}$ &   $117.49^{+1.09}_{-1.08}$ &  $-8.20^{+0.10}_{-0.10}$ &   $1.6 \times 10^5$ \\
      Hercules &     247.7583 &       12.7917 &    $20.6^{+0.2}_{-0.2}$ &    $45.2^{+1.1}_{-1.1}$ &    $-0.03^{+0.04}_{-0.04}$ &    $-0.36^{+0.03}_{-0.03}$ & $131.83^{+12.72}_{-11.60}$ &  $-5.83^{+0.17}_{-0.17}$ &   $1.8 \times 10^4$ \\
 Sagittarius2  &     298.1688 &      $-22.0681$ & $19.32^{+0.03}_{-0.02}$ &  $-177.3^{+1.2}_{-1.2}$ &    $-0.77^{+0.03}_{-0.03}$ &    $-0.89^{+0.02}_{-0.02}$ &    $73.11^{+1.02}_{-0.67}$ &  $-5.20^{+0.10}_{-0.10}$ &   $1.0 \times 10^4$ \\
CanesVenatici2 &     194.2917 &       34.3208 & $21.02^{+0.06}_{-0.06}$ &  $-128.9^{+1.2}_{-1.2}$ &    $-0.15^{+0.07}_{-0.07}$ &    $-0.27^{+0.06}_{-0.06}$ &   $159.96^{+4.48}_{-4.36}$ &  $-5.17^{+0.32}_{-0.32}$ &   $1.0 \times 10^4$ \\
       Tucana5 &     354.3500 &      $-63.2700$ & $18.71^{+0.34}_{-0.34}$ &   $-36.3^{+2.5}_{-2.2}$ &    $-0.39^{+0.03}_{-0.03}$ &    $-0.63^{+0.03}_{-0.03}$ &    $55.21^{+9.36}_{-8.00}$ &  $-5.13^{+0.38}_{-0.38}$ &   $9.6 \times 10^3$ \\
          Leo4 &     173.2375 &       $-0.5333$ & $20.94^{+0.09}_{-0.09}$ &   $132.3^{+1.4}_{-1.4}$ &    $-0.08^{+0.09}_{-0.09}$ &    $-0.21^{+0.08}_{-0.08}$ &   $154.17^{+6.52}_{-6.26}$ &  $-4.99^{+0.26}_{-0.26}$ &   $8.5 \times 10^3$ \\
        Hydra2 &     185.4254 &      $-31.9853$ & $20.64^{+0.16}_{-0.16}$ &   $303.1^{+1.4}_{-1.4}$ &      $-0.34^{+0.1}_{-0.1}$ &    $-0.09^{+0.08}_{-0.09}$ &  $134.28^{+10.27}_{-9.54}$ &  $-4.86^{+0.37}_{-0.37}$ &   $7.5 \times 10^3$ \\
       Hydrus1 &      37.3892 &      $-79.3089$ &  $17.2^{+0.04}_{-0.04}$ &    $80.4^{+0.6}_{-0.6}$ &     $3.79^{+0.01}_{-0.01}$ &     $-1.5^{+0.01}_{-0.01}$ &    $27.54^{+0.51}_{-0.50}$ &  $-4.71^{+0.08}_{-0.08}$ &   $6.5 \times 10^3$ \\
       Carina2 &     114.1067 &      $-57.9992$ & $17.79^{+0.05}_{-0.05}$ &   $477.2^{+1.2}_{-1.2}$ &     $1.88^{+0.01}_{-0.01}$ &     $0.13^{+0.02}_{-0.02}$ &    $36.14^{+0.84}_{-0.82}$ &  $-4.50^{+0.10}_{-0.10}$ &   $5.4 \times 10^3$ \\
    UrsaMajor1 &     158.7200 &       51.9200 &   $19.93^{+0.1}_{-0.1}$ &   $-55.3^{+1.4}_{-1.4}$ &     $1.72^{+0.02}_{-0.02}$ &    $-1.89^{+0.03}_{-0.03}$ &    $96.83^{+4.56}_{-4.36}$ &  $-4.43^{+0.26}_{-0.26}$ &   $5.1 \times 10^3$ \\
    Aquarius2 &     338.4812 &       $-9.3275$ & $20.16^{+0.07}_{-0.07}$ &   $-71.1^{+2.5}_{-2.5}$ &      $-0.17^{+0.1}_{-0.1}$ &    $-0.43^{+0.08}_{-0.08}$ &   $107.65^{+3.53}_{-3.41}$ &  $-4.36^{+0.14}_{-0.14}$ &   $4.7 \times 10^3$ \\
    Leo5 &     172.7900 &        2.2200 & $21.46^{+0.16}_{-0.16}$ &   $173.3^{+3.1}_{-3.1}$ &    $-0.06^{+0.09}_{-0.09}$ &    $-0.25^{+0.09}_{-0.08}$ & $195.88^{+14.98}_{-13.91}$ &  $-4.29^{+0.36}_{-0.36}$ &   $4.4 \times 10^3$ \\
 ComaBerenices &     186.7458 &       23.9042 &    $18.2^{+0.2}_{-0.2}$ &    $98.1^{+0.9}_{-0.9}$ &     $0.41^{+0.02}_{-0.02}$ &    $-1.71^{+0.02}_{-0.02}$ &    $43.65^{+4.21}_{-3.84}$ &  $-4.28^{+0.25}_{-0.25}$ &   $4.4 \times 10^3$ \\
      Columba1 &      82.8600 &      $-28.0300$ &  $21.3^{+0.22}_{-0.22}$ &   $153.7^{+5.0}_{-4.8}$ &     $0.19^{+0.06}_{-0.06}$ &    $-0.36^{+0.06}_{-0.06}$ & $181.97^{+19.40}_{-17.53}$ &  $-4.20^{+0.20}_{-0.20}$ &   $4.1 \times 10^3$ \\
      Pegasus3 &     336.0942 &        5.4200 &   $21.56^{+0.2}_{-0.2}$ &  $-222.9^{+2.6}_{-2.6}$ &       $0.06^{+0.1}_{-0.1}$ &       $-0.2^{+0.1}_{-0.1}$ & $205.12^{+19.79}_{-18.05}$ &  $-4.10^{+0.50}_{-0.50}$ &   $3.7 \times 10^3$ \\
    Reticulum2 &      53.9254 &      $-54.0492$ &    $17.4^{+0.2}_{-0.2}$ &    $64.7^{+1.3}_{-0.8}$ &     $2.39^{+0.01}_{-0.01}$ &    $-1.36^{+0.02}_{-0.02}$ &    $30.20^{+2.91}_{-2.66}$ &  $-3.99^{+0.38}_{-0.38}$ &   $3.4 \times 10^3$ \\
   Triangulum2 &      33.3225 &       36.1783 &    $17.4^{+0.1}_{-0.1}$ &  $-381.7^{+1.1}_{-1.1}$ &      $0.9^{+0.02}_{-0.02}$ &    $-1.26^{+0.02}_{-0.02}$ &    $30.20^{+1.42}_{-1.36}$ &  $-3.90^{+0.20}_{-0.20}$ &   $3.1 \times 10^3$ \\
         Grus2 &     331.0200 &      $-46.4400$ & $18.62^{+0.21}_{-0.21}$ &  $-110.0^{+0.5}_{-0.5}$ &     $0.38^{+0.03}_{-0.03}$ &    $-1.46^{+0.04}_{-0.04}$ &    $52.97^{+5.38}_{-4.88}$ &  $-3.90^{+0.22}_{-0.22}$ &   $3.1 \times 10^3$ \\
   Horologium1 &      43.8821 &      $-54.1189$ &    $19.5^{+0.2}_{-0.2}$ &   $112.8^{+2.5}_{-2.6}$ &     $0.82^{+0.03}_{-0.03}$ &    $-0.61^{+0.03}_{-0.03}$ &    $79.43^{+7.66}_{-6.99}$ &  $-3.76^{+0.56}_{-0.56}$ &   $2.7 \times 10^3$ \\
       Pisces2 &     344.6292 &        5.9525 & $21.31^{+0.17}_{-0.17}$ &  $-226.5^{+2.7}_{-2.7}$ &     $0.16^{+0.08}_{-0.08}$ &        $0.0^{+0.1}_{-0.1}$ & $182.81^{+14.89}_{-13.77}$ &  $-3.67^{+0.60}_{-0.60}$ &   $2.5 \times 10^3$ \\
       Tucana3 &     359.1500 &      $-59.6000$ & $17.01^{+0.16}_{-0.16}$ &  $-102.3^{+2.4}_{-2.4}$ &     $0.54^{+0.06}_{-0.06}$ &    $-1.67^{+0.07}_{-0.07}$ &    $25.23^{+1.93}_{-1.79}$ &  $-3.50^{+0.28}_{-0.28}$ &   $2.1 \times 10^3$ \\
         Grus1 &     344.1767 &      $-50.1633$ &    $20.4^{+0.2}_{-0.2}$ &  $-140.5^{+2.4}_{-1.6}$ &     $0.07^{+0.05}_{-0.05}$ &    $-0.29^{+0.06}_{-0.07}$ & $120.23^{+11.60}_{-10.58}$ &  $-3.47^{+0.59}_{-0.59}$ &   $2.1 \times 10^3$ \\
    Reticulum3 &      56.3600 &      $-60.4500$ & $19.81^{+0.31}_{-0.31}$ &   $274.2^{+7.5}_{-7.4}$ &     $0.36^{+0.14}_{-0.14}$ &     $0.05^{+0.19}_{-0.25}$ &  $91.62^{+14.06}_{-12.19}$ &  $-3.30^{+0.29}_{-0.29}$ &   $1.8 \times 10^3$\\
       Bootes2 &     209.5000 &       12.8500 &  $18.1^{+0.06}_{-0.06}$ &  $-117.0^{+5.2}_{-5.2}$ &    $-2.33^{+0.09}_{-0.08}$ &    $-0.41^{+0.06}_{-0.06}$ &    $41.69^{+1.17}_{-1.14}$ &  $-2.94^{+0.74}_{-0.75}$ &   $1.3 \times 10^3$ \\
      Willman1 &     162.3375 &       51.0500 &    $17.9^{+0.4}_{-0.4}$ &   $-12.3^{+2.5}_{-2.5}$ &     $0.21^{+0.06}_{-0.06}$ &    $-1.08^{+0.09}_{-0.09}$ &    $38.02^{+7.69}_{-6.40}$ &  $-2.90^{+0.74}_{-0.74}$ &   $1.2 \times 10^3$ \\
      Phoenix2 &     354.9975 &      $-54.4061$ &    $19.6^{+0.2}_{-0.2}$ &    $32.4^{+3.7}_{-3.8}$ &     $0.48^{+0.04}_{-0.04}$ &    $-1.17^{+0.05}_{-0.05}$ &    $83.18^{+8.02}_{-7.32}$ &  $-2.70^{+0.40}_{-0.40}$ &   $1.0 \times 10^3$ \\
       Carina3 &     114.6300 &      $-57.8997$ &   $17.22^{+0.1}_{-0.1}$ &   $284.6^{+3.4}_{-3.1}$ &     $3.12^{+0.05}_{-0.05}$ &     $1.54^{+0.06}_{-0.07}$ &    $27.80^{+1.31}_{-1.25}$ &  $-2.40^{+0.20}_{-0.20}$ &   $7.8 \times 10^2$ \\
       Tucana4 &       0.7300 &      $-60.8500$ & $18.41^{+0.19}_{-0.19}$ &    $15.9^{+1.8}_{-1.7}$ &    $-0.14^{+0.06}_{-0.05}$ &    $-1.15^{+0.08}_{-0.06}$ &    $48.08^{+4.40}_{-4.03}$ &  $-1.60^{+0.49}_{-0.49}$ &   $3.7 \times 10^2$ \\
      Sextans1 &     153.2625 &       $-1.6147$ &   $19.67^{+0.1}_{-0.1}$ &   $224.2^{+0.1}_{-0.1}$ &     $0.56^{+0.05}_{-0.05}$ &     $0.07^{+0.06}_{-0.06}$ &    $85.90^{+4.05}_{-3.87}$ &  $-1.60^{+0.76}_{-0.76}$ &   $3.7 \times 10^2$ \\
   Horologium2 &      49.1338 &      $-50.0181$ &   $19.46^{+0.2}_{-0.2}$ & $168.7^{+12.9}_{-12.6}$ &      $0.76^{+0.2}_{-0.29}$ &    $-0.41^{+0.23}_{-0.21}$ &    $77.98^{+7.52}_{-6.86}$ &  $-1.56^{+1.02}_{-1.02}$ &   $3.6 \times 10^2$ \\
       Tucana2 &     342.9796 &      $-58.5689$ &    $18.8^{+0.2}_{-0.2}$ &  $-129.1^{+3.5}_{-3.5}$ &    $-0.08^{+0.01}_{-0.01}$ &    $-1.62^{+0.02}_{-0.02}$ &    $57.54^{+5.55}_{-5.06}$ &  $-1.49^{+0.20}_{-0.20}$ &   $3.4 \times 10^2$ \\
        Segue1 &     151.7667 &       16.0819 &    $16.8^{+0.2}_{-0.2}$ &   $208.5^{+0.9}_{-0.9}$ &    $-2.21^{+0.06}_{-0.06}$ &    $-3.34^{+0.05}_{-0.05}$ &    $22.91^{+2.21}_{-2.02}$ &  $-1.30^{+0.73}_{-0.73}$ &   $2.8 \times 10^2$\\
        Draco2 &     238.1983 &       64.5653 & $16.67^{+0.05}_{-0.05}$ &  $-342.5^{+1.1}_{-1.2}$ &     $1.08^{+0.07}_{-0.07}$ &     $0.91^{+0.08}_{-0.08}$ &    $21.58^{+0.50}_{-0.49}$ &  $-0.80^{+0.40}_{-1.00}$ &   $1.8 \times 10^2$ \\
     UrsaMinor &     227.2854 &       67.2225 &    $19.4^{+0.1}_{-0.1}$ &  $-246.9^{+0.1}_{-0.1}$ &    $-0.32^{+0.14}_{-0.14}$ &     $-0.62^{+0.1}_{-0.08}$ &    $75.86^{+3.58}_{-3.41}$ &  $-0.80^{+0.90}_{-0.90}$ &   $1.8 \times 10^2$ \\
\end{longtable}
Notes: $^a$: no provided error.

\section{Distributions of Predicted vs.~ Actual Halo Masses for Alternate Catalog/Model Pairs}

\label{a:catalog_pairs}

\begin{figure*}
  \begin{minipage}{0.5\textwidth}
    \centering
    \includegraphics[width=\columnwidth]{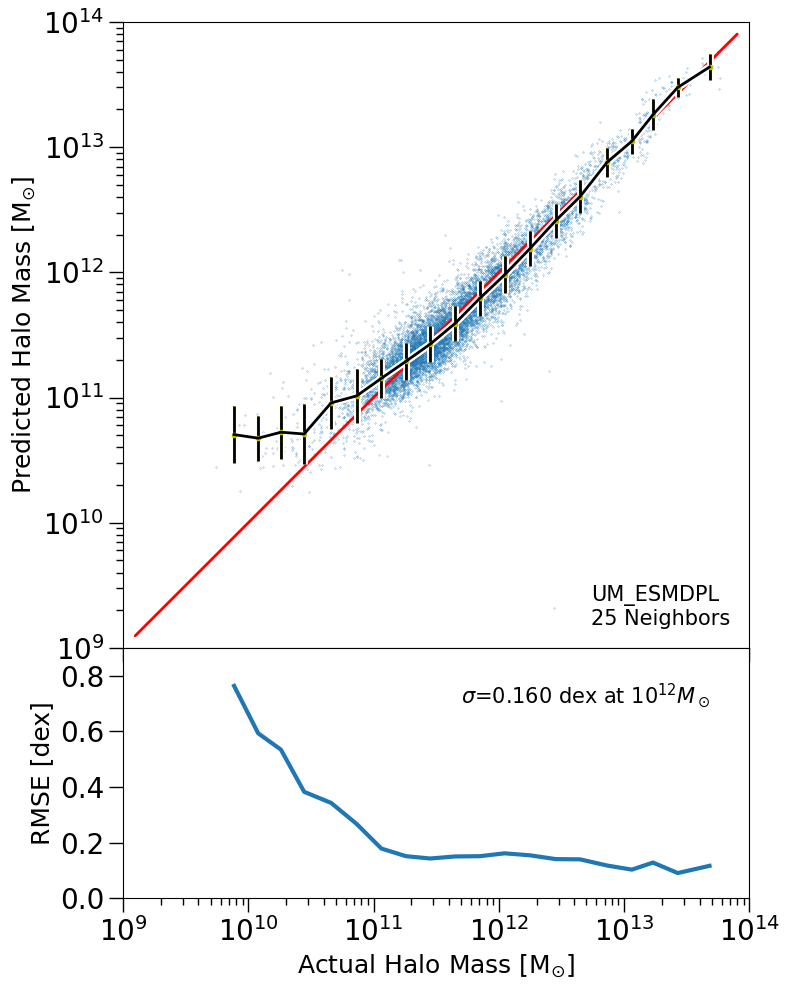}
  \end{minipage}%
  \begin{minipage}{0.5\textwidth}
    \centering
    \includegraphics[width=\columnwidth]{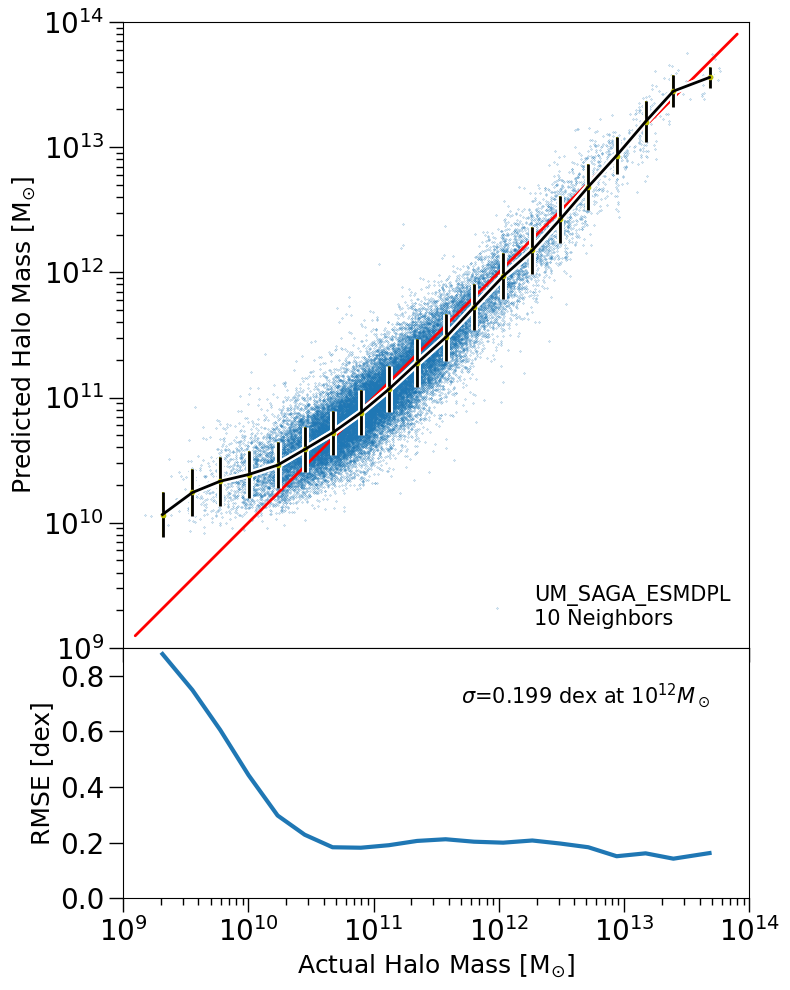}
  \end{minipage}
  \caption{Predicted halo mass versus actual halo mass for the neural networks trained using the UM\_ESMDPL catalog. The left panel shows results for networks using 25 satellites, with an error of 0.160 dex at $10^{12}$ M$_\odot$, while the right panel shows results for 10 satellites, achieving an error of 0.199 dex at $10^{12}$ M$_\odot$. The root mean square error (RMSE) as a function of actual halo mass is displayed in the bottom panels. Error bars indicate the standard deviation of predicted masses in bins of actual halo mass, and the red line represents perfect agreement between predicted and actual masses.}
  \label{fig:result_esmdpl}
\end{figure*}

\begin{figure*}
  \begin{minipage}{0.5\textwidth}
    \centering
    \includegraphics[width=\columnwidth]{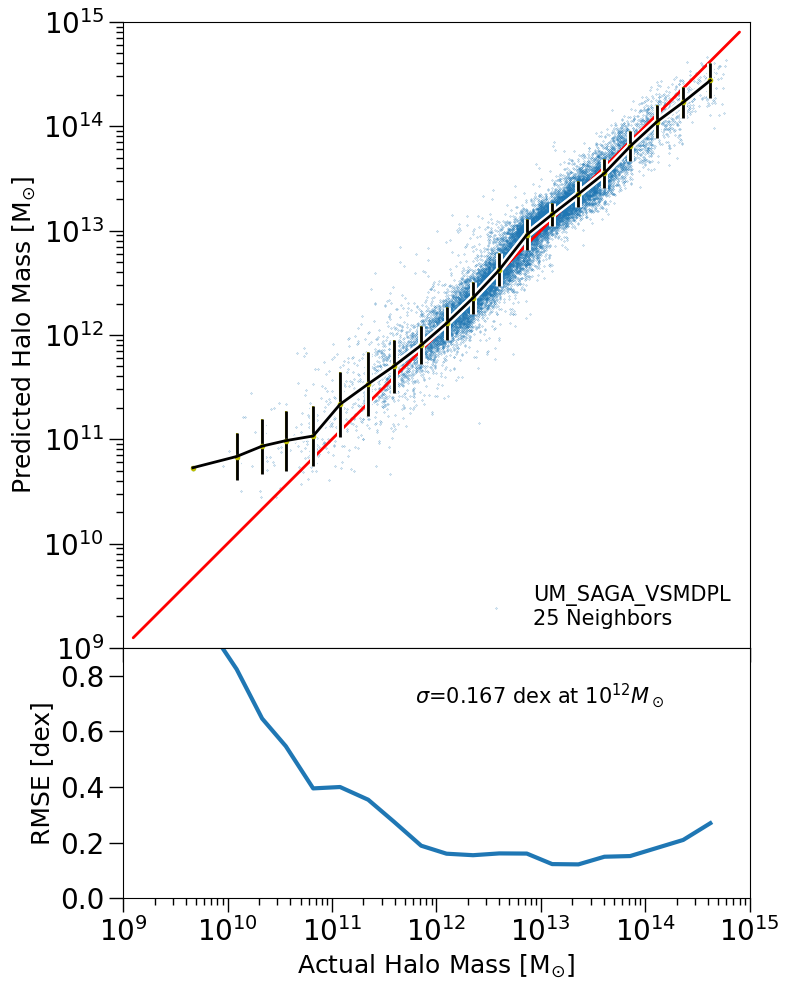}
  \end{minipage}%
  \begin{minipage}{0.5\textwidth}
    \centering
    \includegraphics[width=\columnwidth]{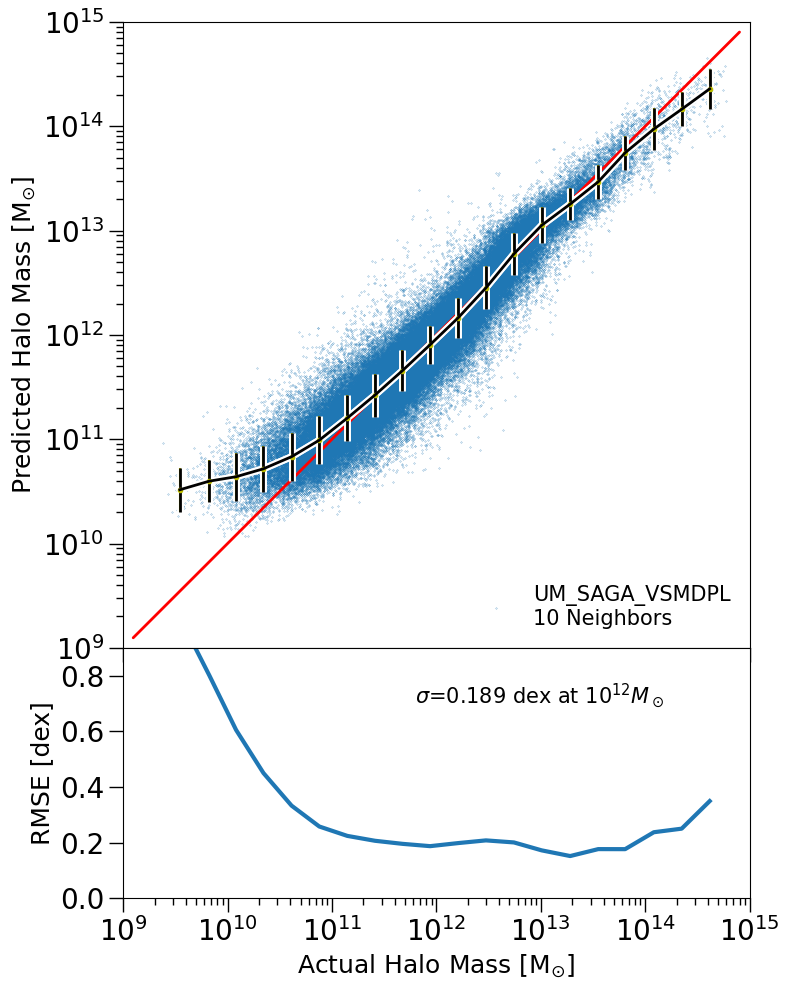}
  \end{minipage}
  \caption{Predicted halo mass versus actual halo mass for the neural networks trained using the UM\_SAGA\_VSMDPL catalog. The left panel shows results for networks using 25 satellites, with an error of 0.167 dex at $10^{12}$ M$_\odot$, while the right panel shows results for 10 satellites, achieving an error of 0.189 dex at $10^{12}$ M$_\odot$. The root mean square error (RMSE) as a function of actual halo mass is displayed in the bottom panels. Error bars show the standard deviation of predicted masses in bins of actual halo mass, and the red line represents perfect agreement between predicted and actual masses.}
  \label{fig:result_SAGA_vsmdpl}
\end{figure*}

\begin{figure*}
  \begin{minipage}{0.5\textwidth}
    \centering
    \includegraphics[width=\columnwidth]{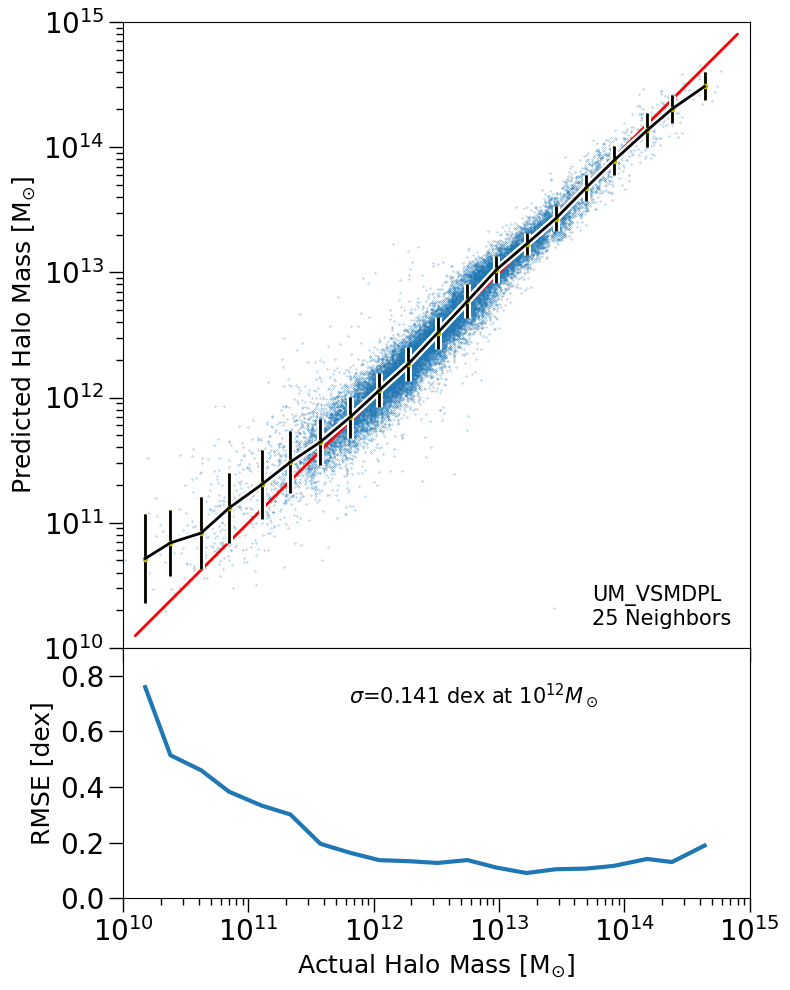}
  \end{minipage}%
  \begin{minipage}{0.5\textwidth}
    \centering
    \includegraphics[width=\columnwidth]{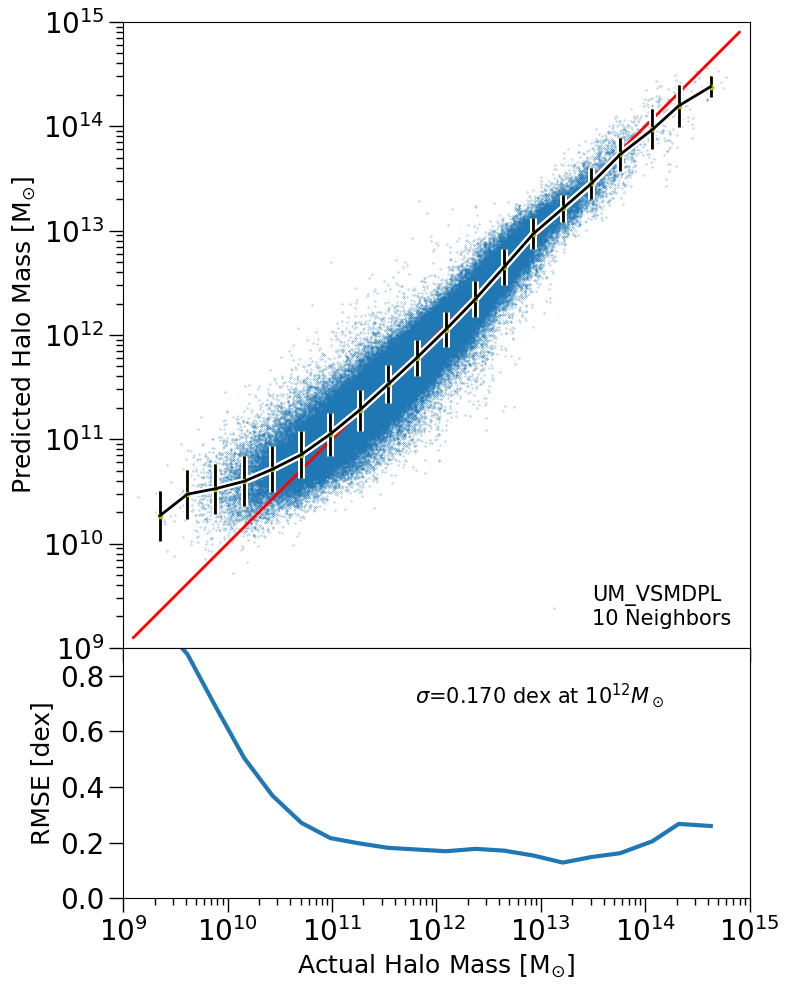}
  \end{minipage}
  \caption{Predicted halo mass versus actual halo mass for the neural networks trained using the UM\_VSMDPL catalog. The left panel shows results for networks using 25 satellites, with an error of 0.141 dex at $10^{12}$ M$_\odot$, while the right panel shows results for 10 satellites, achieving an error of 0.170 dex at $10^{12}$ M$_\odot$. The root mean square error (RMSE) as a function of actual halo mass is displayed in the bottom panels. Error bars show the standard deviation of predicted masses in bins of actual halo mass, and the red line represents perfect agreement between predicted and actual masses.}
  \label{fig:result_vsmdpl}
\end{figure*}

Fig.~\ref{fig:result_esmdpl} shows performance for the ESMDPL/UM simulation/catalog pair, and Figs.~\ref{fig:result_SAGA_vsmdpl} and \ref{fig:result_vsmdpl} show performance for the VSMDPL simulation with the UM-SAGA and UM catalogs, respectively.

\begin{figure*}
  \centering
  \includegraphics[width=\textwidth]{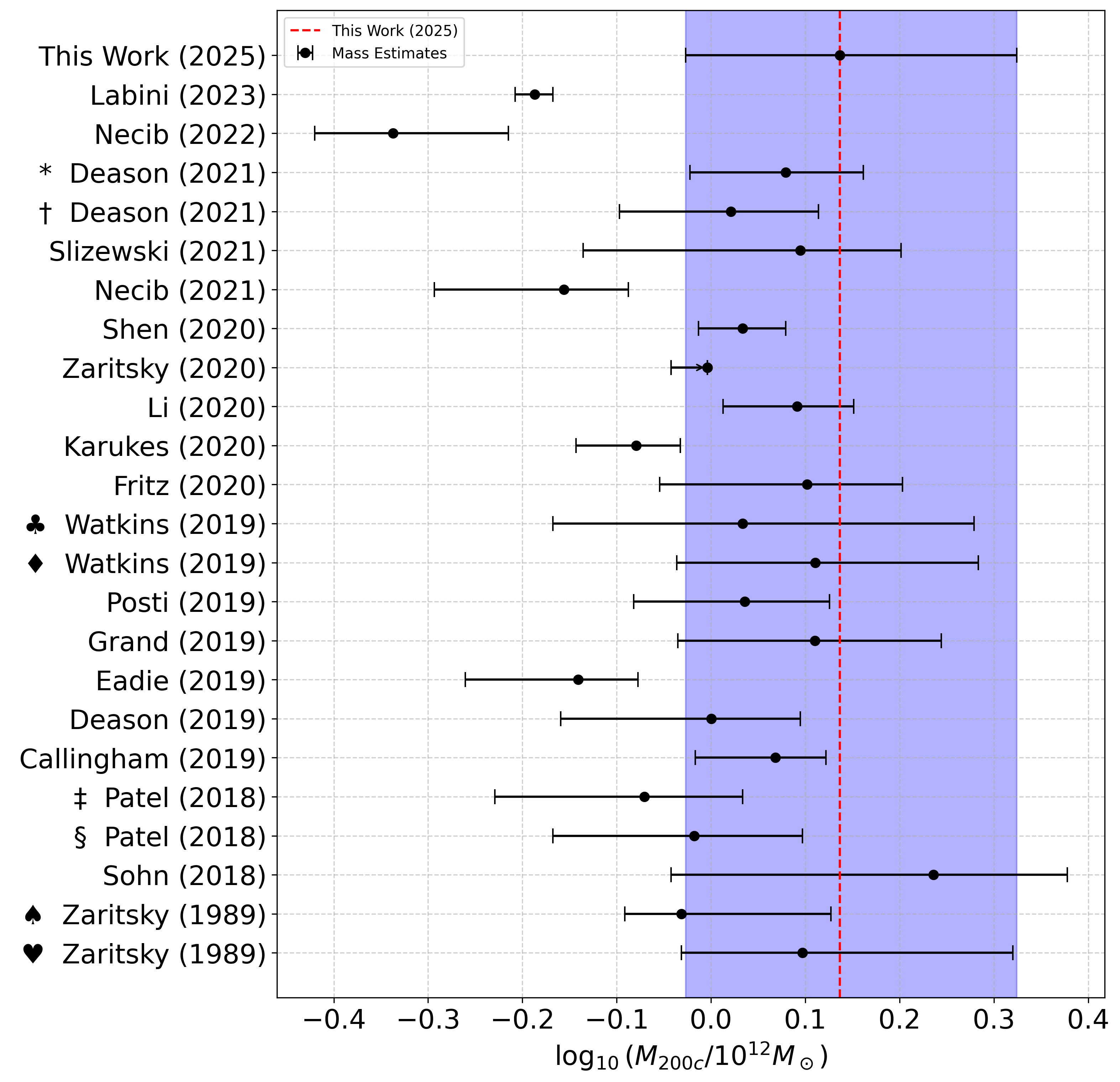}
  \caption{This figure compares Milky Way mass measurements between different studies and this work. All masses are converted to $M_{200c}$. The dashed red vertical line indicates the median of our work, and the shaded band is the $68\%$ credible interval. Measurements are reported in order of year of publication.
  $\ast$Deason et al.~(2021) considers Post-LMC,  
\dag{}Deason et al.~(2021)  considers Pre-LMC,  
 \ddag\citet{Patel18}  includes Sagittarius,  
\S\citet{Patel18} excludes Sagittarius,  
$\clubsuit$ Watkins et al.~(2019) uses \textit{Gaia} Kinematics,  
$\vardiamondsuit$ Watkins et al.~(2019) uses HST,  
$\spadesuit$ Zaritsky et al.~(1989) assumes Radial Orbits,  
$\varheartsuit$ Zaritsky et al.~(1989) assumes Isotropic Orbits.} 
  \label{fig:comparison}
\end{figure*}


\section{What information constrains the mass of the Milky Way?}
\label{a:mw_constraints}

\begin{figure*}
    \begin{minipage}[t]{0.48\textwidth}
        \centering
        \includegraphics[width=\linewidth,valign=c]{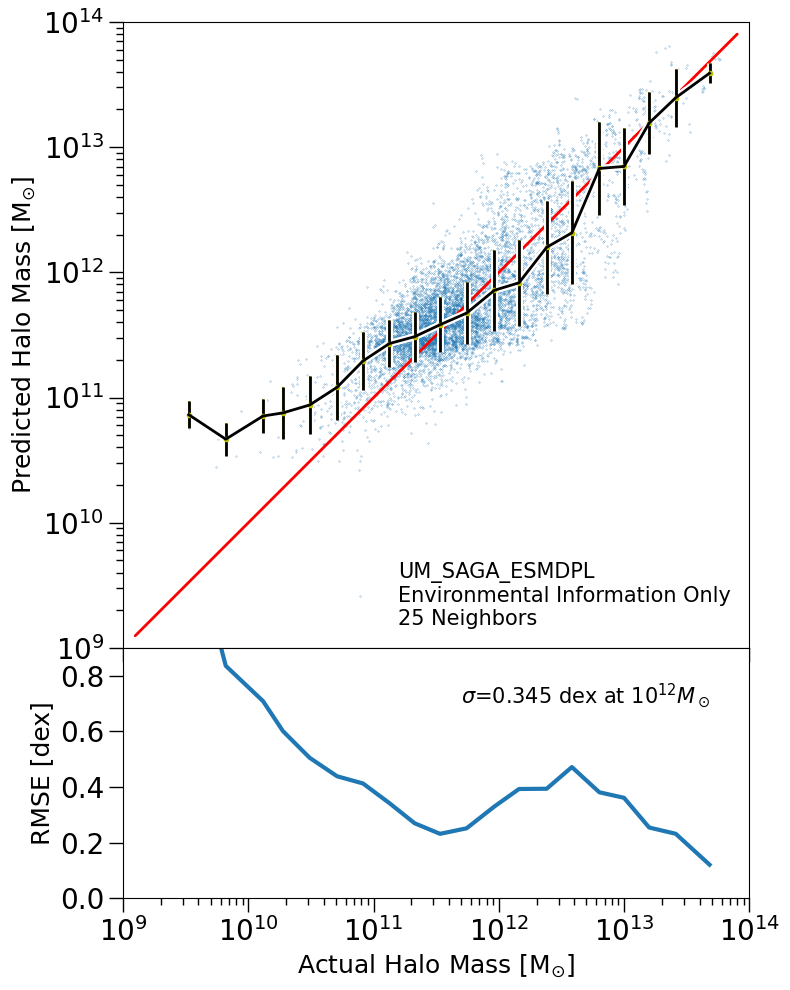}
        \caption{ This figure shows a neural network trained on just three features, the distance to the nearest halo with $M_\mathrm{vir}>10^{14}\Msun$ ($D_{14}$), the distance to the nearest larger halo ($D_\mathrm{larger}$), and the maximum circular velocity of the most massive satellite ($v_\mathrm{max,sat}$). The relatively poor constraints from environmental information suggest that most constraining power arises from the satellite orbits.}
        
        \label{fig:3features}
    \end{minipage}\qquad
    \begin{minipage}[t]{0.48\textwidth}
        \centering
        \includegraphics[width=\linewidth,valign=c]{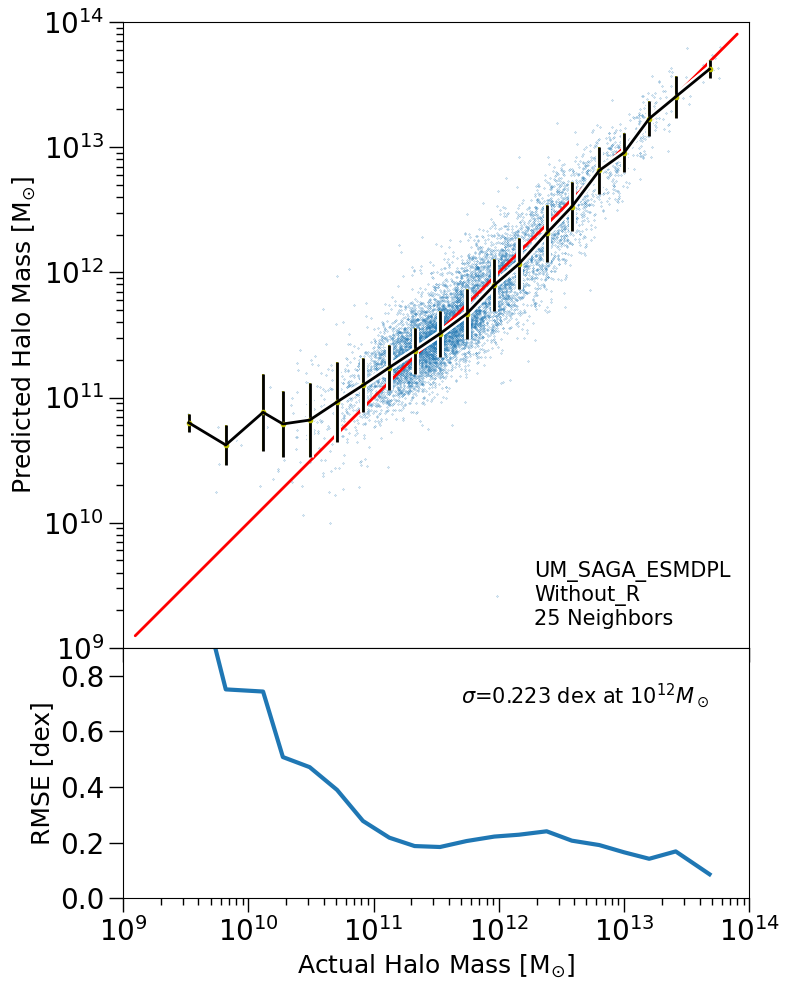}
        \caption{The neural network in this figure was trained on environmental information, plus radial velocity and proper motion, excluding neighbor distance information.}
        \label{fig:without_R}
        \end{minipage}
\end{figure*}

Figure ~\ref{fig:3features} shows that environmental measures (distance to the closest $M_\mathrm{vir}>10^{14}\Msun$ halo [$D_{14}$], distance to the nearest larger halo [$D_\mathrm{larger}$], and $v_\mathrm{max}$ of the most massive satellite [$v_\mathrm{max,sat}$]) provide relatively weak constraints on the host halo mass, with approximately 3$\times$ larger variance than in Fig.\ \ref{fig:histogram}. Unlike $v_\mathrm{max,sat}$ and $D_\mathrm{larger}$, $D_{14}$ does not exhibit much correlation with halo mass \citep{2024OJAp....7E..74H}. This fact suggests that most of the constraining power of the network is coming from the orbits of satellites.

\begin{figure*}
    \begin{minipage}[t]{0.48\textwidth}
        \centering
        \includegraphics[width=\linewidth,valign=c]{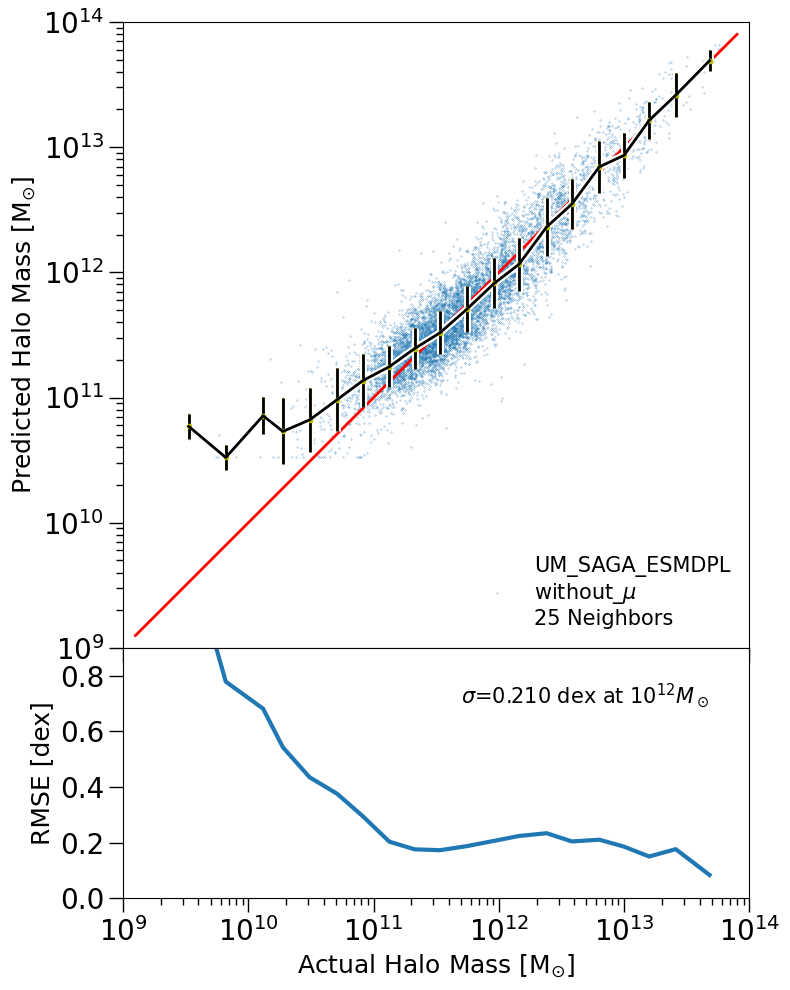}
        \caption{This figure shows a neural network trained on environmental information, distance, and radial velocity, excluding proper motion.}
        
        \label{fig:without_mu}
    \end{minipage}\qquad
    \begin{minipage}[t]{0.48\textwidth}
        \centering
        \includegraphics[width=\linewidth,valign=c]{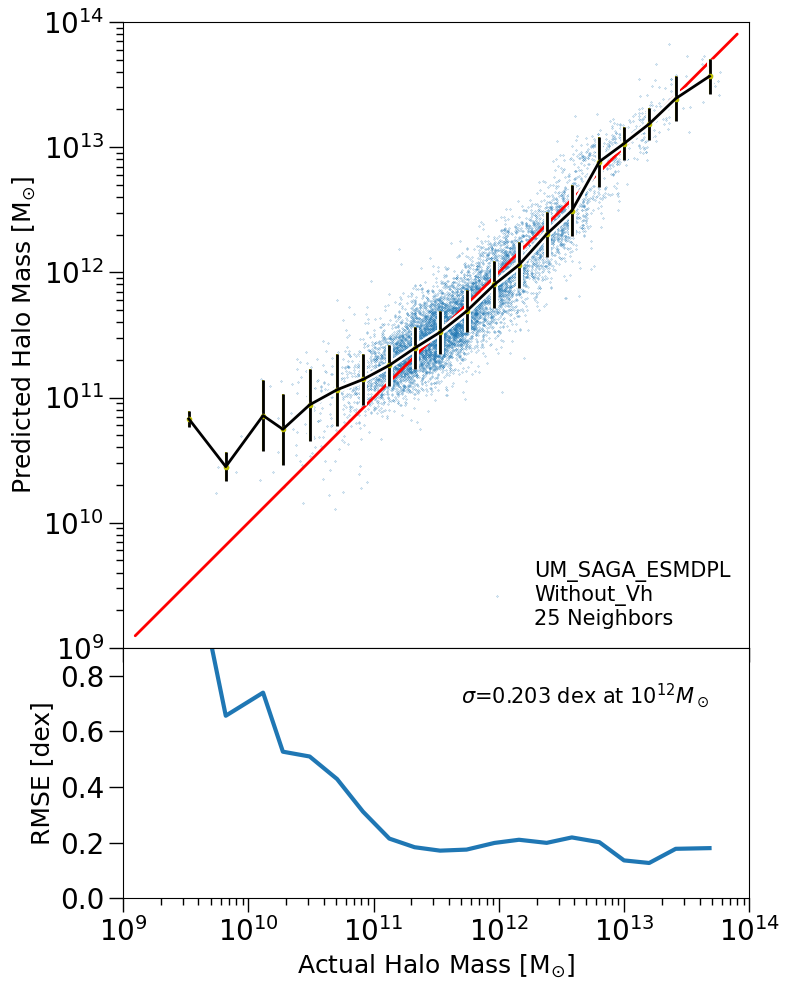}
        \caption{This figure shows a neural network trained on environmental information, distance, and proper motion, excluding radial velocity.}
        \label{fig:without_vh}
        \end{minipage}
\end{figure*}

Figures ~\ref{fig:without_R}, ~\ref{fig:without_mu}, and ~\ref{fig:without_vh} show the effect of removing each of $\mu$, $R$, and $V_\mathrm{los}$ from the input information for satellite orbits.  Removing each one approximately doubles the variance, suggesting that $R$, $V_\mathrm{h}$, and $\mu$ each contribute about equally to the constraints.

As constraints from \textit{Gaia} improve, we expect that reduced observational errors will push the mass constraints closer to those in our previous paper for perfect observations ($\sim 0.12$ dex for 25 neighboring galaxies).  As more satellites are discovered, we expect errors to be reduced, although the scaling with the number of satellites will be less than expected from Poisson statistics due to correlated orbits \citep{2024OJAp....7E..74H}.  We found that going from 10 neighbors to 25 neighbors reduces the errors in recovered halo masses by 30\%, corresponding to uncertainties scaling as $N_\mathrm{neighbors}^{-0.4}$.

\end{document}